\documentclass[iop,twocolappendix]{emulateapj}

\usepackage{natbib}
\usepackage{hyperref}
\usepackage{graphicx}
\usepackage{multirow}
\usepackage{mathtools}

\newcommand{\msun}{\mbox{M}_{\odot}}
\newcommand{\mstar}{$\mbox{M}_*$}

\newcommand{\hii}{H~\textsc{ii}\ }
\newcommand{\ii}{~\textsc{ii}}
\newcommand{\iii}{~\textsc{iii}}

\newcommand{\fdig}{$f_{\text{DIG}}$}

\shorttitle{The MOSDEF survey: a \mstar-SFR-Z relation exists at $\lowercase{z}\sim2.3$}
\shortauthors{Sanders et al.}

\begin{document}

\title{The MOSDEF survey: a stellar mass-SFR-metallicity relation exists at $\lowercase{z}\sim2.3$\altaffilmark{*}}
\altaffiltext{*}{Based on data obtained at the
W.M. Keck Observatory, which is operated as a scientific partnership among the California Institute of Technology, the
University of California, and NASA, and was made possible by the generous financial support of the W.M. Keck Foundation.
}

\author{Ryan L. Sanders\altaffilmark{1}} \altaffiltext{1}{Department of Physics \& Astronomy, University of California, Los Angeles, 430 Portola Plaza, Los Angeles, CA 90095, USA}

\author{Alice E. Shapley\altaffilmark{1}}

\author{Mariska Kriek\altaffilmark{2}} \altaffiltext{2}{Astronomy Department, University of California, Berkeley, CA 94720, USA}

\author{William R. Freeman\altaffilmark{3}} \altaffiltext{3}{Department of Physics \& Astronomy, University of California, Riverside, 900 University Avenue, Riverside, CA 92521, USA}

\author{Naveen A. Reddy\altaffilmark{3,4}} \altaffiltext{4}{Alfred P. Sloan Fellow}

\author{Brian Siana\altaffilmark{3}}

\author{Alison L. Coil\altaffilmark{5}} \altaffiltext{5}{Center for Astrophysics and Space Sciences, University of California, San Diego, 9500 Gilman Dr., La Jolla, CA 92093-0424, USA}

\author{Bahram Mobasher\altaffilmark{3}}

\author{Romeel Dav\'{e}\altaffilmark{6}} \altaffiltext{6}{Institute for Astronomy, Unversity of Edinburgh, James Clerk Maxwell Building, Peter Guthrie Tait Road, Edinburgh, EH9 3FD, United Kingdom}

\author{Irene Shivaei\altaffilmark{7}} \altaffiltext{7}{Department of Astronomy/Steward Observatory, 933 North Cherry Ave, Rm N204, Tucson, AZ, 85721-0065, USA}

\author{Mojegan Azadi\altaffilmark{5}}

\author{Sedona H. Price\altaffilmark{8}} \altaffiltext{8}{Max-Planck-Institut f{\"u}r extraterrestrische Physik, Postfach 1312, Garching, 85741, Germany}

\author{Gene Leung\altaffilmark{5}}

\author{Tara Fetherholf\altaffilmark{3}}

\author{Laura de Groot\altaffilmark{6}} \altaffiltext{6}{Department of Physics, The College of Wooster, 1189 Beall Avenue, Wooster, OH 44691, USA}

\author{Tom Zick\altaffilmark{2}}

\author{Francesca M. Fornasini\altaffilmark{9}} \altaffiltext{9}{Harvard-Smithsonian Center for Astrophysics, 60 Garden Street, Cambridge, MA, 02138, USA}

\author{Guillermo Barro\altaffilmark{10}} \altaffiltext{10}{Department of Phyics, University of the Pacific, 3601 Pacific Ave, Stockton, CA 95211, USA}

\email{email: rlsand@astro.ucla.edu}

\begin{abstract}
We investigate the nature of the relation among stellar mass, star-formation rate, and gas-phase metallicity
 (the \mstar-SFR-Z relation) at high redshifts using a sample of
 260 star-forming galaxies at $z\sim2.3$ from the MOSDEF survey.
  We present an analysis of the high-redshift \mstar-SFR-Z relation based on several
 emission-line ratios for the first time.
  We show that a \mstar-SFR-Z relation clearly exists at $z\sim2.3$.
  The strength of this relation is similar to predictions from
 cosmological hydrodynamical simulations.
  By performing a direct comparison of stacks of $z\sim0$ and $z\sim2.3$
 galaxies, we find that $z\sim2.3$ galaxies have $\sim0.1$~dex lower metallicity
 at fixed \mstar\ and SFR.
  In the context of chemical evolution models,
 this evolution of the \mstar-SFR-Z relation suggests an increase with redshift of the mass-loading factor
 at fixed \mstar, as well as a decrease in the metallicity of infalling gas that is likely due
 to a lower importance of gas recycling relative to accretion from the intergalactic medium
 at high redshifts.
  Performing this analysis simultaneously with multiple metallicity-sensitive line ratios allows
 us to rule out the evolution in physical conditions (e.g., N/O ratio, ionization
 parameter, and hardness of the ionizing spectrum) at fixed metallicity as the source
 of the observed trends with redshift and with SFR at fixed \mstar\ at $z\sim2.3$.
  While this study highlights the promise of performing high-order tests of chemical evolution
 models at high redshifts, detailed quantitative comparisons ultimately await a full
 understanding of the evolution of metallicity calibrations with redshift.
\end{abstract}


\section{Introduction}\label{sec1}

The formation and growth of the dark matter structure of the universe is now well-understood
 because of the relatively simple associated physics and the success of cosmological N-body
 simulations \citep[e.g.,][]{spr05}.
  In contrast, the physics governing the formation and growth of the baryonic content of galaxies
 that inhabit dark matter haloes is more complex.  Thus, the buildup of the gaseous and stellar
 content of galaxies is still not fully understood.
  The chemical enrichment of the interstellar medium (ISM) and its scaling with galaxy properties provides
 a sensitive probe of the key processes governing galaxy growth, including accretion of pristine and
 recycled gas from the intergalactic medium (IGM) and circum-galactic medium (CGM),
 processing of that gas through star formation, and
 feedback from massive stars, supernovae, and AGN that can heat gas and drive outflows.
  The connection between ISM abundance and baryon cycling leads to a
 tight correlation between the stellar masses (\mstar) and
 gas-phase oxygen abundances (12+log(O/H) or Z) of star-forming galaxies
 in the local universe, known as the mass-metallicity relation \citep[MZR; e.g.,][]{tre04,kew08,and13}.
  The MZR has been shown to exist out to $z\sim3.5$, but evolves with redshift such that metallicity at
 fixed \mstar\ decreases with increasing redshift \citep[e.g.,][]{erb06,mai08,tro14,san15,ono16}.

In addition to a fundamental scaling between stellar mass and metallicity, a secondary dependence
 of the $z\sim0$ MZR on star-formation rate (SFR) has been observed \citep[e.g.,][]{ell08,man10,lar10,yat12,and13}.
  The existence of a secondary SFR dependence was first reported by
 \citet{ell08} as a separation in the \mstar-Z plane as a function of specific SFR (sSFR=SFR/\mstar).
  Subsequently, \citet{man10} and \citet{lar10} independently reported a relation among \mstar, metallicity, and SFR
 for $z\sim0$ galaxies and found that the intrinsic scatter around the \mstar-SFR-Z relation is smaller
 than that of the MZR.
  The \mstar-SFR-Z relation is such that at fixed \mstar, galaxies with higher SFRs have lower
 metallicities.  This relation has been interpreted through a theoretical picture
 in which the accretion of pristine or low-metallicity gas from the IGM increases the SFR while
 diluting the metallicity of the ISM.
  At the same time, metallicity increases as the gas reservoir is used
 up in galaxies for which this reservoir is not being replenished.
  \citet{man10} further claim that galaxies up to $z\sim2.5$ lie on the same
 \mstar-SFR-Z relation as $z\sim0$ galaxies, which led these authors to conclude that the
 \mstar-SFR-Z relation is redshift invariant at $z<2.5$.  This non-evolving \mstar-SFR-Z relation
 is known as the ``fundamental metallicity relation'' (FMR).  In the context of the FMR,
 the observed evolution of the MZR with redshift is tied to the evolution of the
 SFR-\mstar\ relation: at earlier epochs, galaxies have higher SFR at fixed \mstar\ on average,
 corresponding to a lower metallicity.  The evolution of SFR at fixed \mstar\ with redshift
 is also tied to an evolution toward higher gas fraction at earlier times \citep{red12,tac13}.

A \mstar-SFR-Z relation is a fundamental feature of galaxy chemical evolution models, including
 the equilibrium model in which the accretion rate is equal to the sum of the SFR and outflow rates
 \citep{fin08,dav12},
 and the gas-regulator model in which the mass of the gas reservoir regulates the SFR that in turn
 determines the outflow rate \citep{lil13}.
  The relationships between inflow rate, outflow rate, and SFR have a similar form at all redshifts
 in these models.
  As such, the theoretical framework through which we understand galaxy growth predicts the existence
 of both the MZR and a \mstar-SFR-Z relation at high redshifts.

When investigating the evolution of the \mstar-SFR-Z relation with redshift, there are two
 questions that must be addressed.
  First, do high-redshift
 galaxies fall on the locally-defined FMR such that the \mstar-SFR-Z relation is redshift invariant?
  Second, do high-redshift galaxies show evidence for the
 existence of a \mstar-SFR-Z relation independent of the $z\sim0$ data?
  It is of interest to confirm SFR dependence of the MZR at high redshifts, in addition to understanding
 the relation between high-redshift galaxies and the $z\sim0$ \mstar-SFR-Z relation.  If the
 metallicities of high-redshift galaxies do not vary with SFR at fixed \mstar, then
 this conflict with current theoretical predictions must be understood in order to obtain
 a more complete picture of galaxy evolution.

There have been many recent studies investigating both of the aforementioned
 questions at $z>1$, but no concensus has been reached about the existence of a high-redshift \mstar-SFR-Z
 relation and its possible redshift invariance.
  Several studies have claimed that galaxies at $z\sim1-3$ are consistent with the $z\sim0$ \mstar-SFR-Z
 relation, suggesting the existence of a redshift invariant \mstar-SFR-Z relation
 \citep{chr12,wuy12,wuy16,bel13,hen13b,sto13,mai14,yab15,hun16a}.
  Many other studies have demonstrated that galaxies at the same redshifts do not follow the
 predictions of the $z\sim0$ FMR, suggesting an evolving \mstar-SFR-Z relation
 \citep{cul14,yab14,zah14b,wuy14,san15,sal15,gra16,kas17}.
  The majority of the literature investigating whether a non-evolving \mstar-SFR-Z relation exists compare
 the predictions of a parameterization of the $z\sim0$ \mstar-SFR-Z relation to high-redshift data,
 and thus often rely on an extrapolation of the $z\sim0$ relation into a SFR regime that few
 if any galaxies in the $z\sim0$ samples occupy.  \citet{san15} and \citet{sal15} showed the
 advantages of instead performing direct, non-parametric comparisons of $z\sim0$ and high-redshift
 galaxies at fixed \mstar, avoiding the uncertainty associated with extrapolations.
  Both studies found that $z\sim2.3$ galaxies have lower metallicities than $z\sim0$ galaxies at
 fixed \mstar\ and SFR.
  It is worth noting that potential redshift evolution of the relations between strong-line ratios and metallicity
 due to evolving gas physical conditions may systematically bias comparisons of metallicity
 over wide redshift ranges and must be carefully considered when investigating the validity of the
 FMR \citep{ste14,ste16,sha15,san16a}.

The question of whether the MZR has a secondary SFR dependence at $z>1$ is naturally more difficult
 to address because it requires detection of a weak secondary effect within the high-redshift sample.
  Several works have searched for a high-redshift \mstar-SFR-Z relation and failed to detect any
 significant SFR dependence within the uncertainties \citep{wuy12,wuy14,wuy16,ste14,yab14,yab15,san15,gra16}.
  While these studies did not observe significant secondary SFR dependence, it is not ruled out due to the
 expected subtlety of the effect.
  The predicted metallicity difference at fixed \mstar\ over the range of SFR probed by such studies
 is small ($\lesssim0.1$~dex), and thus requires either high-precision measurements or large sample sizes to detect.

A handful of studies have claimed to detect a \mstar-SFR-Z relation at high redshifts.
  \citet{sal15} perform a non-parametric comparison between $z\sim0$ galaxies and a sample of
 $\sim130$ $z\sim2.3$ star-forming galaxies.  By comparing metallicity as a function of offset from
 the $z\sim0$ \mstar-sSFR relation, these authors claim to resolve secondary SFR dependence at fixed \mstar,
 albeit with very large scatter and low significance.
  \citet{kac16} divided a sample of 117 $z\sim2.3$ star-forming galaxies at the median SFR, and by fitting
 linear relations to the subsamples found that galaxies with lower SFRs tend to have higher metallicities
 at log($M_*$/M$_{\odot})<10.0$.
  These authors additionally claim that the slope of the $z\sim2.3$ MZR changes with SFR.
  However, roughly half of the low-SFR subsample have only upper limits on the metallicity, and the low-
 and high-SFR subsamples overlap in a narrow \mstar\ range where limits are also present, making it difficult
 to interpret the significance of these results.
  Recently, \citet{kas17} found some evidence for a \mstar-SFR-Z relation at $z\sim1.6$ among individual
 galaxies with metallicity estimates based on [N\ii]$\lambda$6584/H$\alpha$, but do not observe the
 same trends in stacks including non-detections.  The authors conclude that the exclusion of galaxies with non-detections
 of [N\ii] may introduce a bias that falsely makes a \mstar-SFR-Z relation appear.
  \citet{zah14b} presented a detection of the $z\sim1.6$ \mstar-SFR-Z relation, finding a coherent systematic
 shift in the MZR in three \mstar\ bins when dividing their sample of 168 $z\sim1.6$ galaxies at the median SFR.

To date, studies of the high-redshift \mstar-SFR-Z have either been based on
 large samples ($\gtrsim100$) with low S/N individual measurements, or small samples with moderate S/N measurements.
  Previous studies have also largely relied on a single metallicity indicator (most often
 [N\ii]$\lambda$6584/H$\alpha$ or R23=([O\iii]$\lambda\lambda$4959,5007+[O\ii]$\lambda\lambda$3726,3729)/H$\beta$),
 which makes it difficult to ascertain the accuracy of results given the potential redshift evolution
 of various strong-line metallicity calibrations.
  Whether or not a \mstar-SFR-Z relation exists at $z>1$ remains an open question as evidenced by
 the disagreement in the literature.  A clear confirmation of such a relation is needed to test the
 applicability of current galaxy chemical evolution models over cosmic time.

In this paper, we present a study of the $z\sim2.3$ \mstar-SFR-Z relation using data from the
 MOSFIRE Deep Evolution Field (MOSDEF) survey \citep{kri15}.  We present the first analysis of the high-redshift
 \mstar-SFR-Z relation that is based upon multiple metallicity-sensitive emission-line ratios,
 in combination with robust dust-corrected SFRs from measurements of H$\alpha$ and H$\beta$.
  The simultaneous use of multiple metallicity indicators allows us to evaluate whether evolution
 of metallicity calibrations could falsely introduce apparent SFR dependence and therefore provides a more
 robust test of the high-redshift \mstar-SFR-Z relation than previous studies.
  This paper is organized as follows.  We describe the observations, measurements, sample selection,
 and stacking methodology that we employ in Section~\ref{sec2}.
  In Section~\ref{sec3}, we search for a \mstar-SFR-Z relation at $z\sim2.3$ and
 perform a direct comparison to $z\sim0$ galaxies to investigate the redshift evolution of the \mstar-SFR-Z relation.
  We discuss our results and consider the effect of potential biases in Section~\ref{sec4},
 and summarize in Section~\ref{sec5}.  We adopt the following abbrevations for emission-line
 ratios:
\begin{equation}
\text{O}3 = [\mbox{O}~\textsc{iii}]\lambda5007/\text{H}\beta\ ,
\end{equation}
\begin{equation}
\text{N}2 = [\mbox{N}~\textsc{ii}]\lambda6584/\text{H}\alpha\ ,
\end{equation}
\begin{equation}
\text{O3N}2 = \text{O3}/\text{N2}\ ,
\end{equation}
\begin{equation}
\text{N2O2} = [\mbox{N}~\textsc{ii}]\lambda6584/[\text{O}~\textsc{ii}]\lambda\lambda3726,3729\ ,
\end{equation}
\begin{equation}
\text{O}32 = [\mbox{O}~\textsc{iii}]\lambda5007/[\text{O}~\textsc{ii}]\lambda\lambda3726,3729\ ,
\end{equation}
\begin{equation}
\text{R}23 = ([\mbox{O}~\textsc{iii}]\lambda\lambda4959,5007+[\text{O}~\textsc{ii}]\lambda\lambda3726,3729)/\text{H}\beta\ .
\end{equation}
We use the term ``metallicity'' to refer to the gas-phase oxygen abundance (12+log(O/H)) unless
 otherwise specified.  Emission-line wavelengths are given in air.
  We assume a $\Lambda$CDM cosmology with $H_0=70$~km~s$^{-1}$~Mpc$^{-1}$,
 $\Omega_m=0.3$, and $\Omega_{\Lambda}=0.7$.

\section{Observations, Data, and Measurements}\label{sec2}

\subsection{The MOSDEF survey}

The MOSDEF
 survey was a 4-year program in which we utilized the Multi-Object
 Spectrometer For Infra-Red Exploration \citep[MOSFIRE;][]{mcl12} on the 10~m Keck~I telescope
 to obtain near-infrared (rest-frame optical) spectra of galaxies at $1.4\leq z\leq3.8$ \citep{kri15}.
  Galaxies were targeted in three redshift ranges such that strong optical emission lines fall
 in near-infrared windows of atmospheric transmission: $1.37\leq z\leq1.70$, $2.09\leq z\leq2.61$,
 and $2.95\leq z\leq3.80$.  In this study, we focus on the middle redshift range at $z\sim2.3$.
  Targets were selected from the photometric catalogs of the 3D-HST survey
 \citep{bra12,ske14,mom16} based on their redshifts and observed $H$-band (rest-frame optical) magnitudes.
  Spectroscopic or \textit{HST} grism redshifts were used when available; otherwise photometric
 redshifts were utilized for selection.  For the $z\sim2.3$ redshift bin, galaxies were targeted
 down to a fixed $H$-band AB magnitude of $H=24.5$ as measured from \textit{HST}/WFC3 F160W imaging.
  This targeting scheme effectively selects galaxies down to a rough stellar mass limit of
 log($M_*$/M$_{\odot})\sim9.0$, although the sample is not complete below log($M_*$/M$_{\odot})\sim9.5$
 \citep[see][]{shi15}.
  At $2.09\leq z\leq2.61$, [O\ii]$\lambda\lambda$3726,3729 falls
 in the $J$ band; H$\beta$ and [O\iii]$\lambda\lambda$4959,5007 fall in the $H$ band; and
 H$\alpha$, [N\ii]$\lambda\lambda$6548,6584 and [S\ii]$\lambda\lambda$6716,6731 fall in the $K$ band.
  Observations were completed in 2016 May, with the full survey having targeted
 $\sim1500$ galaxies and measured robust redshifts for $\sim1300$ galaxies, with $\sim700$
 at $z\sim2.3$ and $\sim300$ at $z\sim1.5$ and~3.4, respectively.
  Full technical details of the MOSDEF survey observing strategy and data reduction can be found
 in \citet{kri15}.

\subsection{Measurements and derived quantities}

\subsubsection{Stellar mass}

MOSDEF targets have photometric coverage spanning the observed optical to mid-infrared
 (\textit{Spitzer}/IRAC), from which stellar masses can be robustly determined.
  Stellar masses were estimated by fitting the broad- and medium-band photometry from the 3D-HST
 photometric catalogs \citep{ske14,mom16} using the SED
 fitting code FAST \citep{kri09} with the flexible stellar population synthesis models of \citet{con09}.
  A \citet{cha03} initial mass function (IMF), the \citet{cal00} dust reddening curve, solar metallicity, and
 constant star-formation histories were assumed.
  Photometric bands containing significant contamination from high equivalent width
 emission lines (H$\alpha$ in $K$ band and [O\iii]+H$\beta$ in $H$ band) were excluded when fitting the
 SED.
  Uncertainties on the stellar masses were estimated by perturbing the input photometry
 according to the uncertainties and refitting 500 times.

\subsubsection{Emission-line fluxes and redshift}

Emission-line fluxes were determined by fitting Gaussian profiles to the 1D science spectra
 where spectral emission features of interest are expected.  Uncertainties on the emission-line
 fluxes were taken to be the 68th-percentile width of the distribution of line fluxes obtained
 by perturbing the spectrum according to the error spectrum and remeasuring the line fluxes 1000
 times \citep{red15}.  After fitting and subtracting a linear continuum,
 all lines were fit by single Gaussian profiles except for
 [N\ii]$\lambda\lambda$6548,6584+H$\alpha$ and [O\ii]$\lambda\lambda$3726,3729, which were respectively fit by
 triple and double Gaussians simultaneously.  In the case of [O\ii], the width of each component
 of the doublet was required to be identical.  Redshifts were measured from the line with the
 highest S/N ratio for each object, most often H$\alpha$ or [O\iii]$\lambda$5007, and its width
 was used to constrain the width of weaker emission lines.
  Prior to the measurement of line fluxes,
 science spectra were corrected for slit losses on an object-by-object basis, as described in \citet{kri15}.
  Hydrogen Balmer line fluxes were corrected for underlying stellar Balmer absorption by estimating
 the absorption flux from the best-fit SED template, since the stellar continuum is not
 significantly detected for typical MOSDEF targets \citep{red15}.  The Balmer absorption
 corrections were typically $\sim1$\% for H$\alpha$ and $\sim10$\% for H$\beta$.

\subsubsection{Reddening correction and star-formation rate}

SFRs were estimated using dust-corrected H$\alpha$ luminosities by applying
 the calibration of \citet{hao11}, an update to the \citet{ken98} calibration, converted to
 a \citet{cha03} IMF.  Reddening corrections were applied by estimating E(B-V)$_{\text{gas}}$
 from the Balmer decrement (H$\alpha$/H$\beta$) assuming an intrinsic ratio of 2.86 and the
 \citet{car89} Milky Way extinction curve.  Emission-line ratios involving lines
 significantly separated in wavelength were calculated using reddening-corrected line fluxes (N2O2, O32, R23),
 while uncorrected line fluxes were used otherwise (O3N2, N2, O3).
  Uncertainties on both SFRs and reddening-corrected line ratios include the uncertainty in
 E(B-V)$_{\text{gas}}$ in addition to the measurement uncertainties of the line fluxes,
 and SFR uncertainties include an additional 16\% uncertainty added in quadrature to
 account for uncertainty in the slit loss corrections \citep{kri15}.
  With the assumption that the source size is not a strong function of wavelength, the additional
 uncertainty associated with slit loss corrections does not affect the uncertainty of ratios of emission
 lines in the same filter, and does not significantly increase the uncertainty of line ratios with lines
 in multiple filters.

\subsection{Sample selection}

\subsubsection{MOSDEF $z\sim2.3$ sample}

We selected a sample of star-forming galaxies from the full MOSDEF data set with which we
 can investigate the existence of a \mstar-SFR-Z relation at $z\sim2.3$.
  We required a secure redshift determination
 at $2.0\leq z\leq2.7$, log($M_*$/M$_{\odot})\geq9.0$, and a detection of both H$\alpha$ and H$\beta$ at
 signal-to-noise ratio S/N$\geq$3 so that robust estimates of the reddening-corrected SFR may be obtained.
  The stellar mass cut removed six objects with lower stellar masses than the MOSDEF targeting scheme was
 designed to select.  Four of these objects were not main targets but were instead serendipitously detected.
  Both the number of objects targeted and spectroscopic success rate drops precipitously below
 log($M_*$/M$_{\odot})=9.0$ \citep{kri15}.  
  We rejected objects identified
 as AGN by their X-ray or infrared properties \citep{coi15,aza17}, and additionally removed objects
 with [N\ii]$\lambda$6584/H$\alpha>0.5$ when [N\ii]$\lambda$6584 is detected at S/N$\geq$3, which
 are not likely to be dominated by star formation.
  We additionally rejected objects with significant sky-line contamination of H$\alpha$ or
 H$\beta$ such that the line flux is unreliable.
  The requirement that H$\beta$ is cleanly detected at S/N$\geq$3 was the most restrictive of these cuts, removing
 45\% of 462 galaxies that satisfy all other criteria.  We discuss potential biases introduced by the H$\beta$
 detection requirement below.
  This selection yielded a sample of
 260 galaxies at $z_{\text{med}}=2.29$ with stellar masses spanning the range
 log($M_*$/M$_{\odot})=9.0-11.4$ and SFRs ranging from 1.4 to 260~M$_{\odot}$~yr$^{-1}$.
  The median stellar mass and SFR of the sample is log($M_*$/M$_{\odot})=9.92$ and
 22~M$_{\odot}$~yr$^{-1}$, respectively.

\begin{figure*}
 \includegraphics[width=\textwidth]{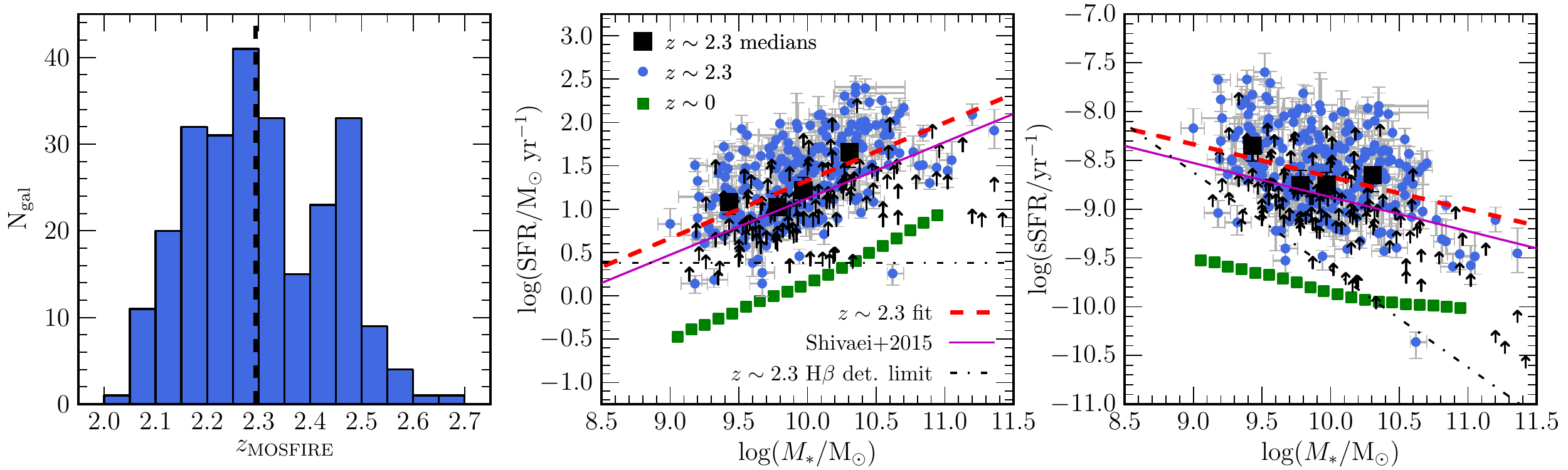}
 \centering
 \caption{Sample properties of the $z\sim2.3$ star-forming galaxy sample from the MOSDEF survey.
  \textsc{Left:} Redshift distribution of the $z\sim2.3$ sample with H$\alpha$ and H$\beta$ detections,
 where the dashed vertical line displays the median redshift of $z_{\text{med}}=2.29$.
  \textsc{Middle:} The SFR-\mstar\ relation for the $z\sim2.3$ sample.  Galaxies at $z\sim2.3$ with
 detections of both H$\alpha$ and H$\beta$ are shown as blue circles with error bars.
  Black squares display the median \mstar\ and SFR of $z\sim2.3$ galaxies with H$\alpha$ and H$\beta$
 detections in four bins of \mstar\ over the range $9.0<\text{log(}M_*/\text{M}_{\odot})<10.5$.
  The best-fit linear relation to the \mstar\ bins is displayed as a red dashed line,
 with best-fit coefficients given in Table~\ref{tab:bestfits}.
  The purple line is the SFR-\mstar\ relation at $1.4<z<2.6$ from \citet{shi15}.
  Black arrows show $3\sigma$ lower limits on SFR and sSFR for galaxies with detected H$\alpha$ but H$\beta$ non-detections.
  The $3\sigma$ H$\beta$ detection threshold at $z\sim2.3$ for MOSDEF observations is indicated
 by the dot-dashed line, scaled from the $5\sigma$ value in \citet{kri15}.
  Green squares denote the \mstar-binned stacks of $z\sim0$ SDSS star-forming galaxies from \citet{and13}.
  \textsc{Right:} The \mstar-sSFR relation for the $z\sim2.3$ sample, with symbols and lines
 the same as in the middle panel.
}\label{fig1}
\end{figure*}

The redshift distribution, and the SFR and specific SFR (sSFR=SFR/$M_*$) as a function of stellar
 mass, of the $z\sim2.3$ sample are shown in Figure~\ref{fig1}.
  Galaxies in this sample scatter about the mean SFR-\mstar\ relation at
 $z\sim2.0-2.5$, with the mean properties falling on the relations
 seen in past studies \citep[e.g.,][]{whi12,spe14,shi15}.
  At log($M_*$/M$_{\odot})<9.5$, galaxies below the mean SFR-\mstar\ relation begin to
 fall below the MOSDEF H$\beta$ detection threshold at $z\sim2.3$ \citep{kri15}.
  \citet{shi15} showed that excluding objects with H$\beta$ non-detections did not
 significantly bias a smaller sample of MOSDEF star-forming galaxies at log($M_*$/M$_{\odot})>9.5$.
  Galaxies for which H$\alpha$ is detected but H$\beta$ is not detected,
 shown as SFR and sSFR lower limits in Figure~\ref{fig1} due to a lack of constraints
 on reddening, are distributed relatively uniformly with \mstar.
  Above log($M_*$/M$_{\odot}$)=9.5, most
 limits lie well above the H$\beta$ detection threshold at $z\sim2.3$, suggesting that
 most of the H$\beta$ non-detections are due to sky-line contamination.
  Since sky-line contamination is essentially a uniform redshift selection that does not
 correlate with other galaxy properties, excluding these H$\beta$ non-detections does
 not bias our sample
 (see Section~\ref{sec:z2msz} for additional discussion of H$\beta$ selection effects).
  \citet{shi15} also demonstrated that below log($M_*$/M$_{\odot})=9.5$, MOSDEF samples
 may be incomplete due to a bias against young objects with small Balmer and 4000~\AA\
 breaks for which the photometric redshifts used in our target selection may be inaccurate.
  This incompleteness does not significantly affect our results because
 the majority of our sample lies above log($M_*$/M$_{\odot})=9.5$.
  The results of our analysis do not significantly change if we restrict the sample to
 log($M_*$/M$_{\odot})>9.5$.

We also investigated whether spectroscopic incompleteness biases our sample.
  Of $\sim700$ galaxies targeted at $2.0\leq z\leq2.7$ in the MOSDEF survey,
 93 failed to produce spectroscopic redshifts.
  Half of these (45/93) are quiescent galaxies based on their rest-frame $UVJ$ colors and thus do not
 affect our sample of star-forming galaxies.  Red star-forming galaxies at $z\sim2.3$ failed to yield redshifts
 15\% of the time, while redshifts were not measured for only 6\% of blue star-forming galaxies.
  These statistics agree with what \citet{kri15} reported based on one-third of the full MOSDEF data set.
  The spectroscopic failures do not occupy any region of $UVJ$ color space in which there are no objects
 with measured redshifts.  Our sample is thus slightly biased against red star-forming galaxies.
  However, our sample is dominated by blue star-forming galaxies ($>$85\%) such that the bias against red
 star-forming galaxies has minimal impact on our results.

\subsubsection{SDSS $z\sim0$ comparison sample}

We compare the high-redshift sample to $z\sim0$ star-forming galaxies using measurements from the composite spectra
 of \citet[][hereafter AM13]{and13}.
  These authors selected a sample of 208,529 star-forming galaxies at $\langle z\rangle=0.078$ from the SDSS \citep{yor00}
 Data Release 7 \citep{aba09} MPA-JHU catalog,\footnote[11]{Available at http://www.mpa-garching.mpg.de/SDSS/DR7/.},
 which includes derived stellar masses \citep{kau03} and aperture-corrected SFR estimates based on
 emission lines \citep{bri04,sal07}.  The spectra of these galaxies were stacked
 in 0.1~dex wide bins in \mstar\ alone, as well as 0.1$\times$0.1~dex bins in both \mstar\ and SFR.
  The \mstar\ and SFR assigned to each stack was taken to be the median stellar
 mass and median SFR of the individual galaxies in each bin.
  We restricted the AM13 stacks to $9.0<\text{log(}M_*/\text{M}_{\odot})<11.0$
 in order to match the stellar mass range in which the majority of the $z\sim2.3$ sample falls.
  We utilize the AM13
 \mstar-only stacks to determine the mean $z\sim0$ relations between \mstar, SFR, sSFR, and
 emission-line ratios, and investigate the scatter about these mean relations using the
 \mstar-SFR stacks.
  The $z\sim0$ comparison sample has the same set of emission lines measured
 as the $z\sim2.3$ sample and SFR measurements based on optical emission lines.
  This similarity allows us to estimate metallicities
 in the same way as for the $z\sim2.3$ sample, and perform a direct comparison of
 line ratios and metallicities at fixed \mstar\ and SFR.

\subsection{Metallicities}\label{sec:metallicities}

The MOSDEF data set crucially provides coverage of several strong rest-optical emission features,
 allowing us to measure multiple metallicity indicators widely applied in the local universe.
  We utilize multiple emission-line ratios to estimate metallicity in this analysis, including
 N2, O3N2, N2O2, and O32.
  We do not utilize R23 to estimate metallicity for reasons described below, but retain the use of this
 metallicity-sensitive line ratio empirically, along with O3.

For N2 and O3N2, we use the empirically calibrated relations of \citet{pet04} based on individual $z\sim0$ \hii regions:
\begin{equation}\label{eq:pp04n2}
12+\log{(\mbox{O/H})}=8.90+0.57\times\mbox{log(N2)} ,
\end{equation}
and
\begin{equation}\label{eq:pp04o3n2}
12+\log{(\mbox{O/H})}=8.73-0.32\times\mbox{log(O3N2)} .
\end{equation}
The intrinsic uncertainties associated with these calibrations are 0.14 and 0.18~dex
 for N2 and O3N2, respectively.  The N2 calibration is valid over the range
 $-2.5<\text{log(N2)}<-0.3$, while the O3N2 calibration is applicable for $\text{log(O3N2)}\lesssim2.0$.
  All of the $z\sim0$ sample and the entire high-redshift MOSDEF sample except for six galaxies
 fall within these bounds.

\citet{bro16} provided a calibration of galaxy metallicity as a function of N2O2 and offset
 from the mean $z\sim0$ sSFR-\mstar\ sequence ($\Delta$log(sSFR)):
\begin{equation}\label{eq:b16n2o2}
12+\log{(\mbox{O/H})}=9.20+0.54\times\mbox{log(N2O2)}-0.36\times\Delta\mbox{log(sSFR)} .
\end{equation}
The $\Delta$log(sSFR) term
 was included because these authors found that N2O2 systematically increases at fixed
 direct-method metallicity as $\Delta$log(sSFR) increases.  \citet{san17} demonstrated that
 the separation of galaxies of different $\Delta$log(sSFR) in the N2O2-metallicity plane
 is a result of contamination of the integrated emission-line fluxes from diffuse ionized gas (DIG),
 which more strongly affects galaxies with lower $\Delta$log(sSFR).
  After correcting the \citet{bro16} strong-line ratios and direct-method metallicities
 for DIG contamination, the dependence on
 $\Delta$log(sSFR) disappears and the galaxies are shifted onto the \hii region metallicity scale.
  To estimate metallicities from N2O2, we perform a linear fit to the corrected \citet{bro16}
 datapoints from \citet{san17}, obtaining the following expression:
\begin{equation}\label{eq:newn2o2}
12+\log{(\mbox{O/H})}=8.94+0.73\times\mbox{log(N2O2)} .
\end{equation}
  The intrinsic scatter around this relation is 0.2~dex, and the calibration is valid over
 the range $-1.3<\text{log(N2O2)}<0.0$.  The entire $z\sim0$ comparison sample falls in this
 range.  Only 10 galaxies in the $z\sim2.3$ sample have $\text{log(N2O2)}<-1.3$, while the
 remainder fall in the range over which the calibration is valid.

We utilize the O32 calibration of \citet{jon15}:
\begin{equation}\label{eq:j15o32}
12+\log{(\mbox{O/H})}=8.3439-0.4640\times\mbox{log(O32)} .
\end{equation}
  The intrinsic calibration uncertainty is 0.11~dex.
  This relation was calibrated using a sample of 113 $z\sim0$ star-forming galaxies with
 metallicity estimates based on electron temperatures, and no evolution in this
 calibration is seen out to $z\sim0.8$ \citep{jon15}.  The calibrating data set only
 covers the range $0.0\lesssim\text{log(O32)}\lesssim1.0$, with the majority of the
 data set lying at $0.0<\text{log(O32)}<0.5$.  While 91 galaxies in
 the $z\sim2.3$ sample and all of the local comparison sample has log(O32)$<$0.0,
 other empirical and theoretical calibrations do not show significant changes to
 the slope of O32 calibrations below log(O32)=0.0 \citep{mai08,cur17}.  We therefore
 apply the \citet{jon15} O32 calibration to galaxies with $-1.0<\text{log(O32)}<1.0$ while
 noting that a systematic bias in metallicity estimates from this calibration
 may be introduced at log(O32)$<$0.0.

While R23 is commonly employed as a metallicity indicator in the local universe, we do not estimate metallicity from R23.
  R23 is double-valued as a function of metallicity, with the turnover occurring at $8.0\lesssim12+\mbox{log(O/H)}\lesssim8.4$
 and $0.7\lesssim\mbox{log(R23)}\lesssim1.0$ \citep[e.g.,][]{mcg91,kew02,pil05}.
  The majority of our $z\sim2.3$ sample lies in this region of parameter space, where small observational uncertainties
 lead to large uncertainties in metallicity and assigning an object to the upper or lower R23 branch is non-trivial.
  Breaking the upper and lower branch degeneracy is usually achieved using N2 or N2O2 in combination with R23
 \citep[e.g.,][]{kew02}.  Star-forming galaxies at $z\sim2.3$ display significant offsets from the $z\sim0$ population
 in strong-line ratio diagrams involving lines of nitrogen \citep{mas14,ste14,sha15,san16a}, however, suggesting that the criterion
 for upper or lower R23 branch assignment based on N2 or N2O2 evolves with redshift.
  Due to these uncertainties in R23-based metallicities for the high-redshift sample, we do not convert R23
 values to metallicities but instead utilize R23 in an empirical sense only.

While all 260 individual galaxies in the $z\sim2.3$ sample have detections of H$\alpha$
 and H$\beta$, some do not have S/N$\geq$3 for [N\ii]$\lambda$6584,
 [O\ii]$\lambda\lambda$3726,3729, and/or [O\iii]$\lambda$5007.
  The sample contains 143 N2 detections, 126 O3N2 detections, 118 N2O2 detections,
 169 O32 and R23 detections, and 223 O3 detections.
  Each of the detected line-ratio subsets has median stellar mass in the range
 $9.9<\text{log(M}_*/\text{M}_{\odot})<10.1$ and median SFR within
 $22<\text{SFR/M}_{\odot}\text{ yr}^{-1}<32$.
  We incorporate information from galaxies with metal line non-detections using a
 stacking analysis as described below.

\subsection{Correcting for diffuse ionized gas contamination}

As mentioned previously, DIG is a significant contaminant of global galaxy
 spectra at $z\sim0$ and introduces systematic biases in galaxy emission-line ratios and
 metallicity estimates \citep{san17}.  In order to obtain accurate metallicities from
 strong-line calibrations, both the measured line ratios and the metallicity calibrations
 must be free of the effects of DIG contamination.  Relations calibrated using individual
 \hii regions such as the \citet{pet04} N2 and O3N2 calibrations are free from this issue.
  We have explicitly corrected the $z\sim0$ sample of \citet{bro16} to the \hii region
 metallicity scale.  While the calibrating data set of the \citet{jon15} O32 relation has
 not been corrected, we do not expect a high level of DIG contamination in this case.
  The galaxies selected by \citet{jon15} are highly star-forming and have high sSFR and
 large H$\alpha$ surface brightness.  \citet{oey07} found that the importance of DIG to
 total galaxy emission decreases with increasing H$\alpha$ surface brightness, such that
 DIG emission is negligible in starburst galaxies.  Using the calibration from \citet{san17}
 based on the \citet{oey07} data set, we estimate that the fraction of total Balmer emission
 originating from DIG (\fdig) is $\lesssim30$\% for the galaxies in the \citet{jon15} sample 
based on their H$\alpha$ surface brightnesses.
  This result suggests that \hii region
 emission dominates the \citet{jon15} calibrating sample and the correction for DIG contamination
 would be minor compared to what is required for typical $z\sim0$ galaxies with \fdig$\approx$55\%
 \citep{san17}.

We corrected the $z\sim0$ comparison sample for DIG contamination using the results of \citet{san17},
 which are based upon empirical models treating galaxies as ensembles of \hii and DIG regions.
  These models also correct for flux-weighting biases present in integrated galaxy measurements because
 the line-emitting regions falling in the spectroscopic aperture have a range of metallicites.
  The AM13 \mstar\ stacks were corrected using the $z\sim0$ model with \fdig=0.55.
  For the AM13 SFR-\mstar\ bins, \citet{san17} determined the median H$\alpha$ surface brightness
 in each bin and constructed a model matched to the corresponding $f_{\text{DIG}}$, since
 the relative importance of DIG varies with SFR.  These corrections allow us to obtain
 more robust metallicity estimates for the $z\sim0$ comparison sample by using the corrected
 line ratios with the metallicity calibrations described above.

There are currently no observational constraints on the importance of DIG at high redshifts.
  However, if high-redshift galaxies follow a similar relation between $f_{\text{DIG}}$
 and H$\alpha$ surface brightness as local galaxies, the high SFRs and compact sizes of
 high-redshift galaxies suggest that DIG emission is negligible at $z\sim2.3$.
  We operate under this assumption, although high-resolution spatially resolved
 emission-line observations of high-redshift galaxies are ultimately required to
 reveal the role of DIG in the early universe.  In addition to compact sizes and high SFRs,
 galaxies have much larger gas fractions at fixed \mstar\ at
 $z\sim2$ than at $z\sim0$ \citep{tac13}, and likely display
 differences in the distribution and importance of different ISM phases.  In light of such
 differences, the assumption that the relation between $f_{\text{DIG}}$ and H$\alpha$
 surface brightness does not change with redshift may need to be revisited.

\subsection{Stacking methodology}

We selected a subset of the sample of 260 $z\sim2.3$ star-forming galaxies with which we perform
 a stacking analysis.  For this stacking subsample, we additionally required that the spectrum
 of each object had wavelength coverage of [O\ii]$\lambda\lambda$3726,3729, H$\beta$,
 [O\iii]$\lambda$5007, H$\alpha$, and [N\ii]$\lambda$6584, and that the stellar mass
 fell in the range $9.0\leq\log(M_*$/M$_{\odot})\leq10.5$.  The wavelength coverage criterion
 allows for the measurement of all emission-line ratios of interest in this study from the
 stacked spectra.  The stellar mass criterion is designed to only stack over the mass range
 in which the MOSDEF $z\sim2.3$ star-forming sample is not obviously incomplete based on target selection
 and/or spectroscopic success rate.  \citet{kri15} showed that both the number of galaxies
 targeted and the spectroscopic success rate sharply decrease at log($M_*$/M$_{\odot})<9.0$
 using data from the first 2 years of the MOSDEF survey.  Such behavior is expected because
 of the target selection down to fixed $H$-band magnitude and the inherent faintness of low-mass
 targets.  A drop in the spectroscopic success rate at log($M_*$/M$_{\odot})>10.5$ was also
 identified and found to be partially caused by a significantly lower success rate for red
 star-forming and quiescent galaxies compared to blue star-forming galaxies \citep{kri15}.
  We have confirmed these trends in success rate with the full MOSDEF $z\sim2.3$ sample.
  Applying the above criteria yields
 a stacking subsample of 195 galaxies at $z_{\text{med}}=2.29$ with median stellar mass
 log($M_*$/M$_{\odot})=9.89$ and median SFR of 18~M$_{\odot}$~yr$^{-1}$.
  The stacking subsample has slightly lower median stellar mass and SFR than the larger $z\sim2.3$
 sample it was selected from, primarily because of the upper stellar mass cut.

We used two different binning methods for this analysis.  We binned in stellar mass in order to
 determine mean $z\sim2.3$ line-ratio and metallicity relations as a function of stellar mass.
  Galaxies were stacked in four stellar mass bins selected such that an equal number of galaxies
 fell in each bin (48-49 galaxies per bin).  We refer to the stacks binned in stellar mass
 only as the ``\mstar\ stacks.''
  We also binned in both stellar mass and offset from the mean $z\sim2.3$ sSFR-\mstar\ relation ($\Delta$log(sSFR))
 in order to assess the presence of a \mstar-SFR-Z relation at $z\sim2.3$ and its evolution from
 $z\sim0$.  We found that dividing the stacking sample at $\Delta$log(sSFR/yr$^{-1})=\pm0.2$ splits the
 sample into three roughly equal parts, with 68 galaxies in the highly star-forming subsample
 with $\Delta$log(sSFR$)>0.2$, 67 galaxies falling within 0.2~dex of the mean relation, and
 60 galaxies with $\Delta$log(sSFR$)<-0.2$.  Since these bins were selected with respect to
 the mean $z\sim2.3$ sSFR-\mstar\ relation, the stellar mass distribution in each bin is
 similar except for the $\Delta$log(sSFR$)<-0.2$ bin, in which objects with
 log($M_*$/M$_{\odot})\lesssim9.5$ fall below the MOSDEF H$\beta$ detection limit at $z\sim2.3$.
  The $\Delta$log(sSFR$)<-0.2$ bin is thus biased toward higher \mstar.
  After dividing the sample in $\Delta$log(sSFR), galaxies in each $\Delta$log(sSFR) bin were
 stacked in two stellar mass bins divided at the median stellar mass, yielding a total of 6
 \mstar-$\Delta$log(sSFR) bins.
  We refer to the stacks binned in both \mstar\ and $\Delta$log(sSFR) as the ``\mstar-$\Delta$sSFR stacks.''

We produced composite spectra by first shifting each spectrum into the rest frame and
 converting from flux density to luminosity density using the spectroscopic redshift.
  Since some of the line ratios of interest require reddening correction, we corrected
 each individual spectrum for reddening by applying a reddening correction to the luminosity
 density at each wavelength element based on the Balmer decrement (H$\alpha$/H$\beta$) of each galaxy,
 assuming the \citet{car89} extinction curve.
  This process effectively only corrects the nebular lines, since the continuum is not
 significantly detected for individual galaxies in our stacking sample.  Each reddening-corrected spectrum
 was then normalized by the dust-corrected H$\alpha$ luminosity.  Spectra were interpolated onto
 a wavelength grid with spacing equal to the wavelength sampling at the median redshift of
 the sample, equal to 0.40~\AA\ in $J$ band, 0.50~\AA\ in $H$ band, and 0.66~\AA\ in $K$ band.
  Normalized composite spectra were created by taking the median value of the normalized spectra at each
 wavelength increment, and multiplying by the median dust-corrected H$\alpha$ luminosity to produce
 the final composite spectrum in units of luminosity density.

Emission-line luminosities were measured from stacked spectra by fitting a flat continuum to regions
 around emission features and Gaussian profiles to the emission lines following the fitting
 methodology for individual spectra.  Uncertainties on emission-line fluxes were estimated
 using a Monte Carlo technique in which we perturbed the stellar masses according to the
 uncertainties and divided the perturbed sample into
 the same stellar mass bins as before, bootstrap resampled each bin to account for sample
 variance, perturbed the individual spectra according to the error spectra, perturbed the
 individual E(B-V)$_{\text{gas}}$ values and applied them to correct the individual spectra for
 reddening, stacked according to the method described above,
 and remeasured the emission-line luminosities.  This process was repeated 100 times, and the
 uncertainties on line luminosities were obtained from the 68th-percentile width of the
 distribution of remeasured line luminosities.  The median Balmer absorption luminosity of
 the individual galaxies in each bin was utilized to apply Balmer absorption corrections
 to the stacked H$\alpha$ and H$\beta$ luminosities.

Emission-line ratios and uncertainties
 were calculated using the line luminosities and uncertainties measured from the stacked spectra.
  We found that the stacking method described above
 robustly reproduces line ratios characteristic of the individual galaxies in a stack by comparing
 the stacked line ratios to median line ratios of individual galaxies with significant
 detections of all of the emission lines of interest.
  Stellar mass and SFR were assigned to each stack using the median \mstar\ and median SFR of the
 individual galaxies in each bin.  Uncertainties on the median \mstar\ and SFR were estimated using
 the distribution of median values produced by the Monte Carlo simulations.
  The median \mstar\ and SFR of galaxies in each bin and
 measured line ratios of the \mstar\ and \mstar-$\Delta$sSFR stacks are presented in Table~\ref{tab:stacks}.


\begin{table*}[t]
 \caption{Galaxy properties and emission-line ratios from stacks of $z\sim2.3$ star-forming galaxy spectra.
 }\label{tab:stacks}
 \begin{tabular*}{\textwidth}{@{\extracolsep{\fill}} c r r r r r r r r r }
   \hline\hline
   N$_{\mbox{gal}}$\tablenotemark{a} & $\log{\left(\frac{\mbox{M}_*}{\mbox{M}_\odot}\right)}$\tablenotemark{b} & SFR$_{\mbox{med}}$\tablenotemark{c} & log$\left(\frac{\mbox{sSFR}}{\mbox{yr}^{-1}}\right)$\tablenotemark{d} & log(N2) & log(O3) & log(O3N2) & log(N2O2) & log(O32) & log(R23) \\
   \hline
   \hline
   \multicolumn{10}{c}{M$_*$ stacks} \\
   \hline
   49 & 9.0-9.62; $9.43^{+0.01}_{-0.03}$ & $12.1^{+3.4}_{-1.2}$ & $-8.34^{+0.16}_{-0.05}$ & $-1.22^{+0.10}_{-0.05}$ & $0.66^{+0.04}_{-0.02}$ & $1.89^{+0.06}_{-0.10}$ & $-1.13^{+0.20}_{-0.08}$ & $0.28^{+0.10}_{-0.13}$ & $0.93^{+0.04}_{-0.03}$ \\
   49 & 9.62-9.89; $9.78^{+0.02}_{-0.01}$ & $10.6^{+0.8}_{-4.4}$ & $-8.75^{+0.03}_{-0.15}$ & $-1.08^{+0.06}_{-0.09}$ & $0.55^{+0.05}_{-0.02}$ & $1.63^{+0.11}_{-0.07}$ & $-1.12^{+0.09}_{-0.18}$ & $0.07^{+0.16}_{-0.11}$ & $0.89^{+0.06}_{-0.05}$ \\
   49 & 9.89-10.13; $9.97^{+0.01}_{-0.04}$ & $16.6^{+2.8}_{-6.8}$ & $-8.75^{+0.11}_{-0.13}$ & $-0.90^{+0.05}_{-0.06}$ & $0.45^{+0.02}_{-0.04}$ & $1.36^{+0.06}_{-0.06}$ & $-0.93^{+0.13}_{-0.14}$ & $-0.05^{+0.07}_{-0.17}$ & $0.84^{+0.05}_{-0.05}$ \\
   48 & 10.14-10.50; $10.30^{+0.01}_{-0.02}$ & $45.1^{+14.9}_{-9.3}$ & $-8.65^{+0.17}_{-0.07}$ & $-0.71^{+0.03}_{-0.03}$ & $0.37^{+0.02}_{-0.05}$ & $1.09^{+0.03}_{-0.06}$ & $-0.80^{+0.21}_{-0.15}$ & $-0.20^{+0.23}_{-0.19}$ & $0.83^{+0.07}_{-0.16}$ \\
   \hline
   \multicolumn{10}{c}{M$_*$-$\Delta$sSFR stacks: $\Delta$log(sSFR/yr$^{-1})<-0.2$} \\
   \hline
   30 & 9.18-9.90; $9.72^{+0.04}_{-0.01}$ & $5.7^{+0.2}_{-1.5}$ & $-8.96^{+0.01}_{-0.12}$ & $-1.05^{+0.10}_{-0.11}$ & $0.43^{+0.04}_{-0.04}$ & $1.49^{+0.13}_{-0.14}$ & $-1.04^{+0.13}_{-0.16}$ & $0.00^{+0.08}_{-0.11}$ & $0.80^{+0.05}_{-0.04}$ \\
   30 & 9.90-10.49; $10.04^{+0.02}_{-0.05}$ & $11.1^{+1.9}_{-0.6}$ & $-8.99^{+0.10}_{-0.01}$ & $-0.79^{+0.09}_{-0.06}$ & $0.34^{+0.04}_{-0.05}$ & $1.14^{+0.06}_{-0.10}$ & $-0.75^{+0.13}_{-0.09}$ & $-0.11^{+0.04}_{-0.12}$ & $0.76^{+0.05}_{-0.04}$ \\
   \hline
   \multicolumn{10}{c}{M$_*$-$\Delta$sSFR stacks: $-0.2\leq\Delta\log$(sSFR/yr$^{-1})\leq+0.2$} \\
   \hline
   34 & 9.00-9.86; $9.50^{+0.01}_{-0.03}$ & $9.2^{+0.4}_{-2.1}$ & $-8.53^{+0.03}_{-0.07}$ & $-1.16^{+0.04}_{-0.17}$ & $0.59^{+0.03}_{-0.04}$ & $1.76^{+0.14}_{-0.07}$ & $-1.09^{+0.10}_{-0.19}$ & $0.21^{+0.09}_{-0.13}$ & $0.88^{+0.03}_{-0.05}$ \\
   33 & 9.86-10.47; $10.1^{+0.05}_{-0.01}$ & $26.2^{+4.4}_{-0.1}$ & $-8.68^{+0.07}_{-0.03}$ & $-0.81^{+0.07}_{-0.04}$ & $0.41^{+0.04}_{-0.04}$ & $1.23^{+0.05}_{-0.08}$ & $-0.83^{+0.11}_{-0.08}$ & $-0.11^{+0.04}_{-0.09}$ & $0.84^{+0.05}_{-0.02}$ \\
   \hline
   \multicolumn{10}{c}{M$_*$-$\Delta$sSFR stacks: $\Delta$log(sSFR/yr$^{-1})>+0.2$} \\
   \hline
   34 & 9.20-9.92; $9.55^{+0.01}_{-0.05}$ & $32.1^{+3.0}_{-4.1}$ & $-8.04^{+0.07}_{-0.04}$ & $-1.23^{+0.07}_{-0.10}$ & $0.69^{+0.03}_{-0.01}$ & $1.93^{+0.11}_{-0.08}$ & $-1.22^{+0.11}_{-0.11}$ & $0.25^{+0.15}_{-0.06}$ & $0.97^{+0.03}_{-0.04}$ \\
   34 & 9.92-10.5; $10.25^{+0.01}_{-0.03}$ & $88.2^{+17.3}_{-1.0}$ & $-8.30^{+0.12}_{-0.02}$ & $-0.78^{+0.05}_{-0.05}$ & $0.46^{+0.03}_{-0.03}$ & $1.24^{+0.08}_{-0.05}$ & $-0.97^{+0.07}_{-0.07}$ & $-0.18^{+0.09}_{-0.06}$ & $0.92^{+0.03}_{-0.04}$ \\
   \hline
 \end{tabular*}
 \tablenotetext{1}{Number of galaxies in a bin.}
 \tablenotetext{2}{Range and median $\log{(\mbox{M}_*/\msun)}$ of galaxies in a bin.}
 \tablenotetext{3}{Median dust-corrected H$\alpha$ SFR of galaxies in a bin.}
 \tablenotetext{4}{The sSFR assigned to each stack is calculated using the median SFR and median \mstar\ of galaxies in each bin.}
\end{table*}

\section{Results}\label{sec3}

A sensitive test of secondary dependences in the MZR is to look for correlations
 between deviations from the mean MZR and deviations from mean relations of other galaxy properties
 as a function of stellar mass, such as SFR, or H~\textsc{i} and H$_2$ gas fraction.
  In particular, a signature of a \mstar-SFR-Z relation is an
 anticorrelation between residuals about the mean MZR ($\Delta$log(O/H)) and residuals about
 the mean sSFR-\mstar\ relation ($\Delta$log(sSFR)).
  We search for this signature of a \mstar-SFR-Z relation at $z\sim2.3$.
  We first fit the mean sSFR-\mstar\ relation, as well as the mean relations between \mstar\
 and the emission-line ratios O3N2, N2, N2O2, O32, O3, and R23.  We additionally fit the mean MZR
 based on metallicities determined using each of the line ratios O3N2, N2, N2O2, and O32.
  We then examine the relationship between the residuals in line ratios (and the corresponding
 metallicity values) and sSFR relative to the mean relations.
  A similar method was employed by \citet{sal15} and \citet{kas17}.

While we use sSFR residuals in order to make a comparison to the simulation predictions of
 \citet{dav17}, we note that using residuals around the mean SFR-\mstar\ relation would give the
 same result since an offset of a galaxy above or below the mean SFR-\mstar\ relation in log(SFR)
 is equal to the offset of that galaxy from the mean sSFR-\mstar\ relation in log(sSFR).
  Accordingly, the slope of the $\Delta$log(O/H) vs. $\Delta$log(sSFR) relation provides a
 direct probe of the dependence of O/H on SFR at fixed \mstar.

\subsection{Mean $z\sim2.3$ relations}

The \mstar\ stacks are shown in the SFR-\mstar\ and sSFR-\mstar\ planes in Figure~\ref{fig1}.
We determine the mean $z\sim2.3$ sSFR-\mstar\ relation by fitting a linear function
 to the \mstar\ stacks in the sSFR-\mstar\ plane.
The best-fit coefficients are listed in Table~\ref{tab:bestfits}.
  We determine the residuals about the mean \mstar-sSFR relation
 by subtracting this mean relation from the sSFR estimate for each
 galaxy in the $z\sim2.3$ sample.  
  In Figure~\ref{fig1}, we compare our mean relation to that obtained by \citet{shi15} using the first 2 years of
 MOSDEF data.  The two relations are generally consistent, although we find a slightly
 higher normalization, likely due in part to different SED fitting assumptions and the resulting
 stellar masses.
  We note that the MOSDEF $z\sim2.3$ H$\beta$ detection limit biases the SFR in the lowest-mass
 bin high, and thus the fit we derive here may be steeper in the sSFR-\mstar\ plane and shallower in the SFR-\mstar\
 plane than the true mean relation of the $z\sim2.3$ star-forming population.  However, our results do not
 change significantly if we instead fit mean relations excluding the lowest-mass bin.
  Incompleteness in the lowest-mass bin due to the H$\beta$ detection limit thus does not
 affect our conclusions.

\begin{table}[t]
 \centering
 \caption{Best-fit linear coefficients to $z\sim2.3$ galaxy properties as a function of stellar mass.}\label{tab:bestfits}
 \begin{tabular}{ c | r r }
   \hline\hline
   property & slope & intercept \\ \hline\hline
   star formation\tablenotemark{a} & & \\ \hline
   SFR & 0.67 & $-5.33$ \\
   sSFR & $-0.33$ & $-5.33$ \\ \hline\hline
   line ratios\tablenotemark{a} & & \\ \hline
   O3N2 & $-0.94$ & 10.72 \\
   N2 & 0.59 & $-6.82$ \\
   N2O2 & 0.41 & $-5.03$ \\
   O32 & $-0.56$ & 5.51 \\
   O3 & $-0.34$ & 3.90 \\
   R23 & $-0.12$ & 2.03 \\ \hline\hline
   12+log(O/H)\tablenotemark{b} & & \\ \hline
   O3N2 & 0.30 & 5.30 \\
   N2 & 0.34 & 5.01 \\
   N2O2 & 0.30 & 5.27 \\
   O32 & 0.26 & 5.79 \\
   \hline\hline
 \end{tabular}
 \tablenotetext{1}{Coefficients for $\log(X)=m\times\log(M_*/\text{M}_{\odot})+b$, where $X$ is the appropriate galaxy property, $m$ is the slope, and $b$ is the intercept.}
 \tablenotetext{2}{Coefficients for $12+\log(\text{O/H})=m\times\log(M_*/\text{M}_{\odot})+b$, where $m$ is the slope, $b$ is the intercept, and 12+log(O/H) is determined using the corresponding line ratio.}
\end{table}

\begin{figure}
 \includegraphics[width=\columnwidth]{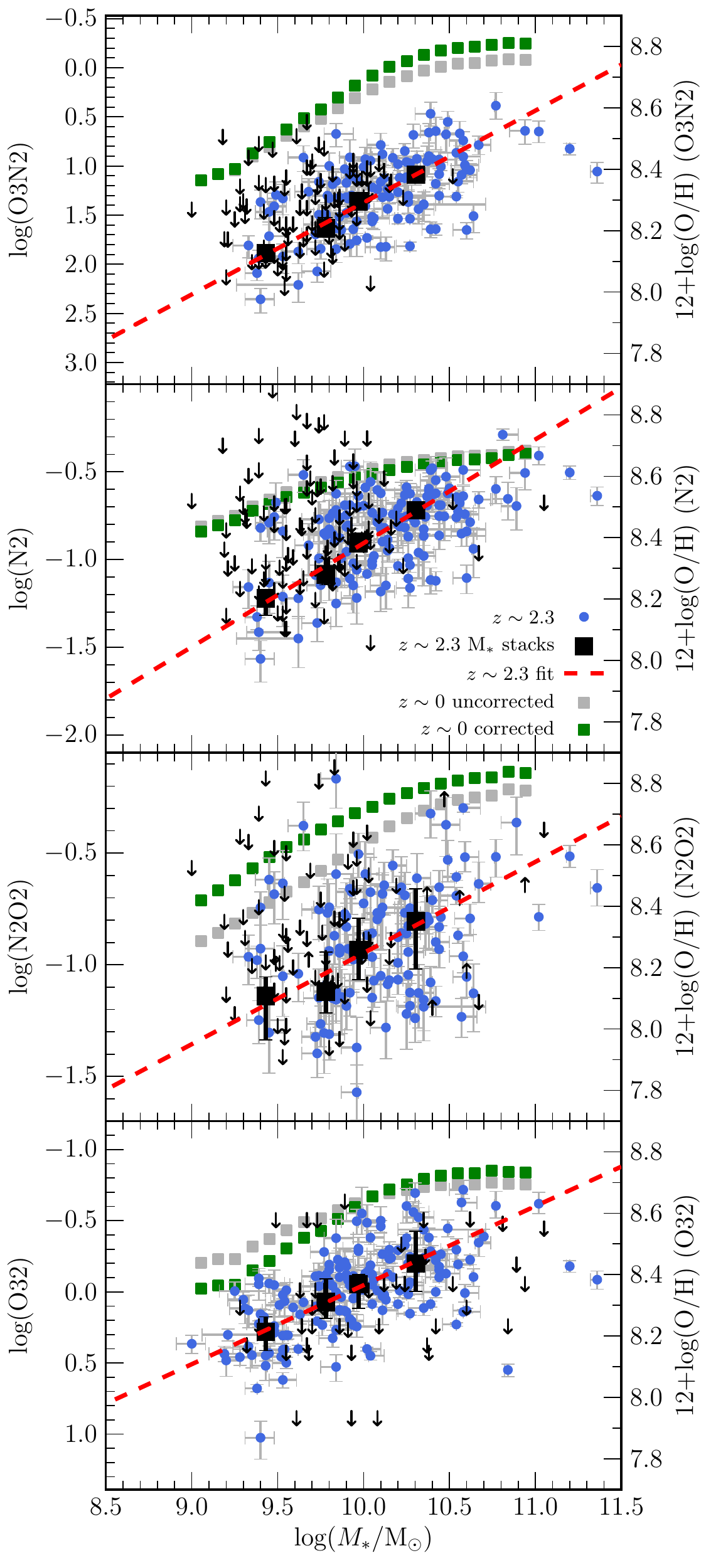}
 \centering
 \caption{The emission-line ratios O3N2, N2, N2O2, and O32 vs.~\mstar\ for the $z\sim2.3$ and $z\sim0$ comparison
 samples.  The left y-axis displays the emission-line ratio values, while the right y-axis displays the
 translation into metallicity according to equations~\ref{eq:pp04n2}, \ref{eq:pp04o3n2}, \ref{eq:j15o32},
 and~\ref{eq:newn2o2}.  All y-axes are oriented such that metallicity increases upwards, and span the same
 range in metallicity.  The $z\sim2.3$ galaxies with SFR measurements are shown as blue circles when all
 of the required emission lines are detected for each line ratio, while black arrows display $3\sigma$ limits
 when one or more of the required lines is not detected.  Stacks of both detections and non-detections in bins
 of \mstar\ are indicated by black squares.  The red dashed line is the best-fit linear relation to the stacks
 in each panel, with the best-fit parameters given in Table~\ref{tab:bestfits}.
  Gray squares show the $z\sim0$ \mstar\ stacks of \citet{and13} prior to correction for DIG contamination,
 while green squares show the same stacks after correcting for contamination from DIG using the models of
 \citet{san17}.
}\label{fig2}
\end{figure}

In Figure~\ref{fig2}, we present the excitation- and
 metallicity-sensitive line ratios O3N2, N2, N2O2, and O32 vs. \mstar.  The line ratios
 O3N2 and O32 are sensitive to the ionization parameter, containing both a high and low ionization
 energy ionic species.  N2 is also sensitive to changes in ionization parameter, as well as the
 nitrogen abundance (N/H), such that higher ionization parameter and lower N/H lead to lower N2.
  In contrast, N2O2 is primarily sensitive to changes in the N/O abundance ratio \citep{kew02}.
  We observe a clear progression toward higher ionization parameter
 and lower N/O and N/H at fixed stellar mass from $z\sim0$ to $z\sim2.3$.  These results are in
 agreement with what has been found for other samples at $z>2$ \citep[e.g.,][]{erb06,ste14,sha15,hol16}.
  While recent results for high-redshift samples have suggested the possibility of redshift evolution
 of physical conditions such as the ionizing spectrum, ionization parameter, or N/O at fixed O/H
 \citep[e.g.,][]{kew13,mas14,mas16,ste14,ste16,sha15,san16a,str17},
 the observation of both higher ionization parameter and lower
 nitrogen abundance at fixed \mstar\ is difficult to explain without chemical evolution
 (i.e., lower O/H at fixed \mstar\ with increasing redshift) playing a major role.
  We determine the mean relation between these strong-line ratios and \mstar\ by fitting
 linear functions to the \mstar\ stacks, and present the best-fit coefficients in
 Table~\ref{tab:bestfits}.

All four line ratios indirectly trace oxygen abundance, since
 ionization parameter is tightly anticorrelated with metallicity \citep{dop06a,dop06b,per14,sanc15},
 while N/O and N/H are
 correlated with metallicity due to the secondary production channel of nitrogen \citep{kew02,pet04,per09}.
  The observation of such correlations in the local universe
 has led to the construction of metallicity calibrations based on these strong-line ratios.
  The translation of the strong-line ratios O3N2, N2, N2O2, and O32 into metallicity is shown by the
 right set of y-axes in Figure~\ref{fig2}.
  All four panels are matched to the same range in metallicity (7.7$<$12+log(O/H)$<$8.9).
  Under the assumption
 that metallicity calibrations do not strongly evolve with redshift, the presence of higher excitation
 and lower N/O at $z\sim2.3$ corresponds to lower metallicity at fixed \mstar\
 compared to $z\sim0$.  We find that $z\sim2.3$ galaxies are offset toward lower metallicities at fixed \mstar by
 0.37, 0.25, 0.46, and 0.25~dex on average for metallicities based on O3N2, N2, N2O2, and O32,
 respectively.  The best-fit linear MZR for these indicators can be found by passing the
 line-ratio vs. \mstar\ fits through the metallicity calibrations
 (equations~\ref{eq:pp04n2}, \ref{eq:pp04o3n2}, \ref{eq:newn2o2}, and~\ref{eq:j15o32}).
  We list the best-fit coefficients for the MZR in Table~\ref{tab:bestfits}, and calculate
 O/H residuals relative to these best-fit linear relations.

The ability of such locally calibrated relations to accurately estimate nebular metallicities
 in high-redshift galaxies has been called into question by recent studies
 \citep[e.g.,][]{ste14,san15,sha15,ste16,str17}.
  We note here that the method we employ to search for a \mstar-SFR-Z relation
 at $z\sim2$ using
 residuals around mean relations is immune to changes in the normalizations of metallicity
 calibrations since we are looking at \textit{relative} differences in metallicity,
 but would be affected by changes in the slopes.
  The interpretation of our results could also be affected by changes in ionized gas physical
 conditions that correlate with sSFR at fixed \mstar.
  We discuss these potential systematic effects on our results in Section~\ref{sec:evolution}.

\subsection{Is there SFR dependence of the $z\sim2.3$ MZR?}\label{sec:z2msz}

As previously mentioned, while it is of interest to determine whether or not $z\sim0$
 and high-redshift galaxies lie on the same \mstar-SFR-Z relation
 (i.e., the FMR holds out to high redshifts), it is more fundamental to
 establish whether the metallicities of high-redshift galaxies display a secondary
 dependence on SFR at fixed \mstar.  The existence of a high-redshift \mstar-SFR-Z
 relation is a prerequisite for interpreting the position of high-redshift galaxies
 relative to the FMR observed at $z\sim0$.  We look for a \mstar-SFR-Z relation at
 $z\sim2.3$ by searching for correlated scatter around the MZR and \mstar-sSFR relation.

\begin{figure}
 \includegraphics[width=\columnwidth]{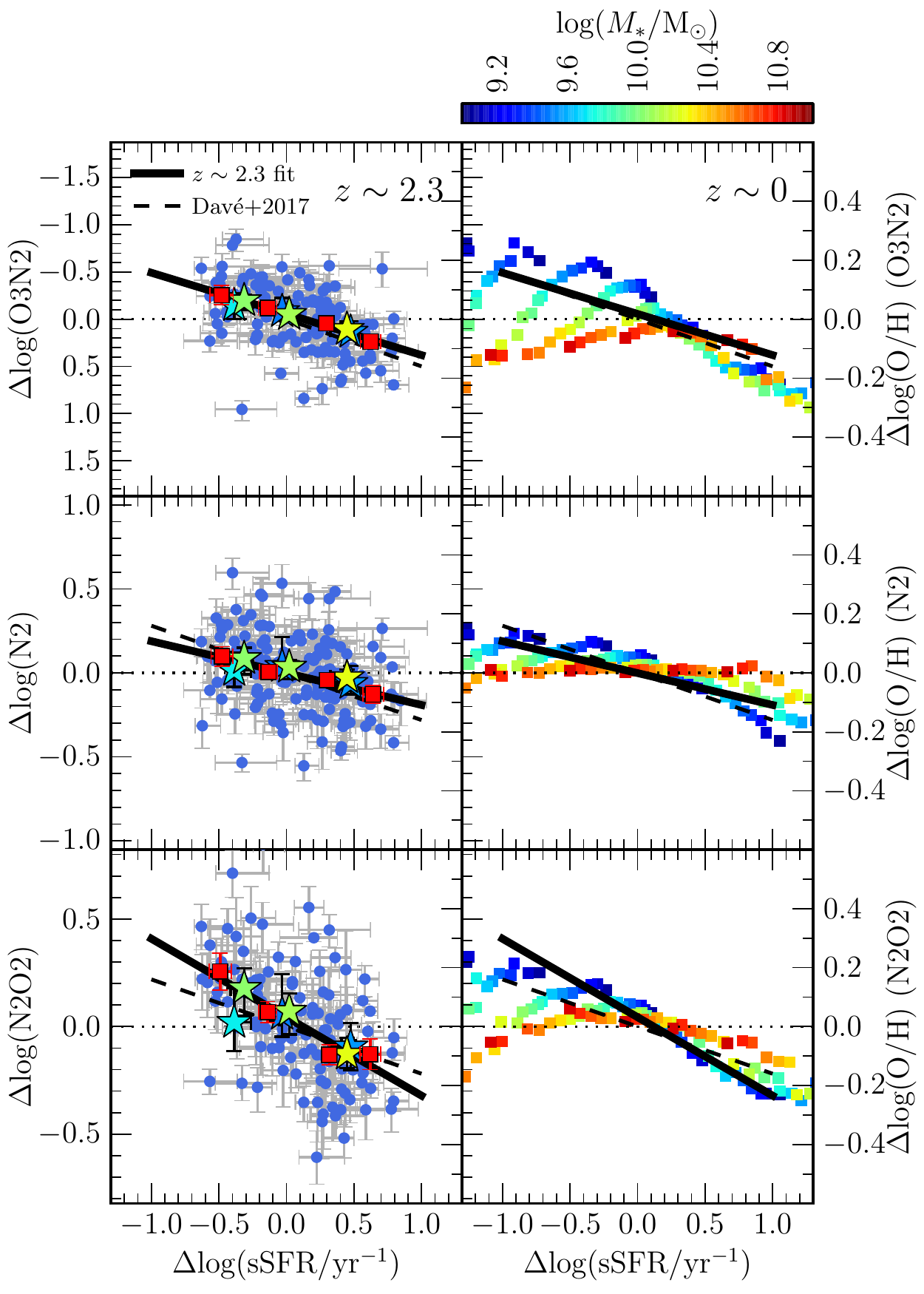}
 \centering
 \caption{Deviation plots comparing the residuals around mean line ratio and metallicity vs.~\mstar\
 relations ($\Delta$log(line ratio) and $\Delta$log(O/H)) to residuals around the mean
 \mstar-sSFR relation ($\Delta$log(sSFR/yr$^{-1}$)).  The top, middle, and bottom rows display
 residuals for line ratio and metallicities based upon O3N2, N2, and N2O2, respectively.
  In each panel, the left vertical axis shows the scale of the residuals in each line ratio,
 while the right vertical axis displays the scale of the corresponding residuals in metallicity.
  The $z\sim2.3$ sample is shown in the left column, while the right column presents
 residuals for the $z\sim0$ \mstar-SFR stacks of \citet{and13}, color-coded by \mstar.
  In the left column,
 blue circles denote individual $z\sim2.3$ galaxies, red squares show medians in bins of
 $\Delta$log(sSFR/yr$^{-1}$), and colored stars display the $z\sim2.3$ \mstar-$\Delta$sSFR stacks,
 color-coded by \mstar.
  The solid black line shows the best fit to the $z\sim2.3$ medians,
 with best-fit slope given in Table~\ref{tab:slopes}.
  The prediction from the cosmological simulations of \citet{dav17} is presented as
 the black long-dashed line.
}\label{fig3}
\end{figure}

In the left column of Figure~\ref{fig3}, we show the $z\sim2.3$ residuals around the
 mean \mstar\ relations for the line ratios O3N2, N2, and N2O2, and the corresponding O/H residuals,
 all as a function of residuals around the $z\sim2.3$ mean \mstar-sSFR relation.
  We show both $z\sim2.3$ individual galaxies and the \mstar-$\Delta$sSFR stacks,
 the latter color-coded by stellar mass.
  There is a clear trend among both the stacks and individual galaxies such that
 the sSFR residuals correlate with O3N2 residuals and anticorrelate with residuals in N2 and N2O2.
  These trends are as expected in the presence of a \mstar-SFR-Z relation, given that O3N2 decreases
 with increasing metallicity while N2 and N2O2 increase.
  Furthermore, the correlations are statistically significant based on Spearman correlation tests.
  The correlation coefficients and $p$-values are shown in Table~\ref{tab:slopes}.
  The corresponding trends in $\Delta$log(O/H) are similar for all three metallicity indicators:
 $\Delta$log(O/H) decreases with increasing $\Delta$log(sSFR).  This trend in the residuals demonstrates
 that $z\sim2.3$ galaxies with higher SFR have lower O/H at fixed stellar mass,
 confirming the existence of a \mstar-SFR-Z relation at $z\sim2.3$.  This is the first time that such
 a relation has been clearly demonstrated to exist at this redshift.

We quantify the strength of the $z\sim2.3$ \mstar-SFR-Z relation by
 fitting linear relations to the median $\Delta$log(O/H) in bins of $\Delta$log(sSFR).
  The best-fit lines are shown in the middle column of Figure~\ref{fig3},
 and the best-fit slopes are given in Table~\ref{tab:slopes}.
  These slopes are inconsistent with a flat relation (i.e., no dependence of O/H on SFR at fixed \mstar)
 at $3-4\sigma$.  The \mstar-$\Delta$sSFR stacks fall on the same relations as the medians of the
 individual $z\sim2.3$ galaxies.  The high-mass \mstar-$\Delta$sSFR stacks most clearly follow the same
 relation as the medians of the individual points, while the low-mass \mstar-$\Delta$sSFR stacks are noisier
 but all fall within 1$\sigma$ of the best-fit relations.  The position of the stacks relative
 to the individual detections suggests that galaxies with non-detections in these
 emission-line ratios lie on the same anticorrelations.  We conclude that requiring line-ratio
 detections in the individual galaxies does not significantly bias these results.

\begin{table}[t]
 \centering
 \caption{Best-fit slopes and correlation tests of $\Delta$log(O/H) vs. $\Delta$log(sSFR/yr$^{-1}$), where O/H is estimated using the indicated strong-line ratio.}\label{tab:slopes}
 \begin{tabular}{ c l l l }
   \hline\hline
   line ratio & slope & $r_\text{s}$\tablenotemark{a} & $p$-value\tablenotemark{b} \\ \hline
   O3N2 & $-0.14\pm0.034$ & 0.48 & $1.5\times10^{-8}$ \\
   N2 & $-0.11\pm0.037$ & $-0.32$ & $8.6\times10^{-5}$ \\
   N2O2 & $-0.27\pm0.067$ & $-0.54$ & $3.2\times10^{-10}$ \\
   O32 & $-0.037\pm0.022$ & 0.14 & 0.07 \\
   \hline\hline
 \end{tabular}
 \tablenotetext{1}{Spearmen correlation coefficient.}
 \tablenotetext{2}{Probability of the sample being drawn from an uncorrelated distribution.}
\end{table}

The right column of Figure~\ref{fig3} displays the line-ratio and O/H residuals as a function of
 $\Delta$log(sSFR) for the $z\sim0$ \mstar-SFR stacks, color-coded by \mstar.
  In each panel, we include the best-fit line to the $z\sim2.3$ sample for comparison.
  The $z\sim0$ stacks show similar trends at log($M_*$/M$_{\odot})\lesssim10.0$,
 but display the well-known decrease in the strength and disappearance of the
 SFR dependence at high stellar masses \citep{man10,yat12,sal14}, 
 as evidenced by the tendency of high-mass stacks to lie closer to $\Delta$log(O/H$)=0$ at fixed
 $\Delta$log(sSFR/yr$^{-1})$.
  A weakening of the $z\sim0$ sSFR dependence at $\Delta$log(sSFR/yr$^{-1})<0$ is also apparent
 in all three indicators, as pointed out by \citet{sal14}.
  There is no evidence for such a flattening of the \mstar-SFR-Z relation
 at high masses or $\Delta$log(sSFR/yr$^{-1})<0$ among the $z\sim2.3$ \mstar-$\Delta$sSFR stacks.
  However, we note that the mass range in which the $z\sim0$ sample displays little to no SFR dependence
 (log($M_*$/M$_{\odot})\gtrsim10.5$) lies above the mass range probed by the $z\sim2.3$ sample.

\begin{figure}
 \includegraphics[width=\columnwidth]{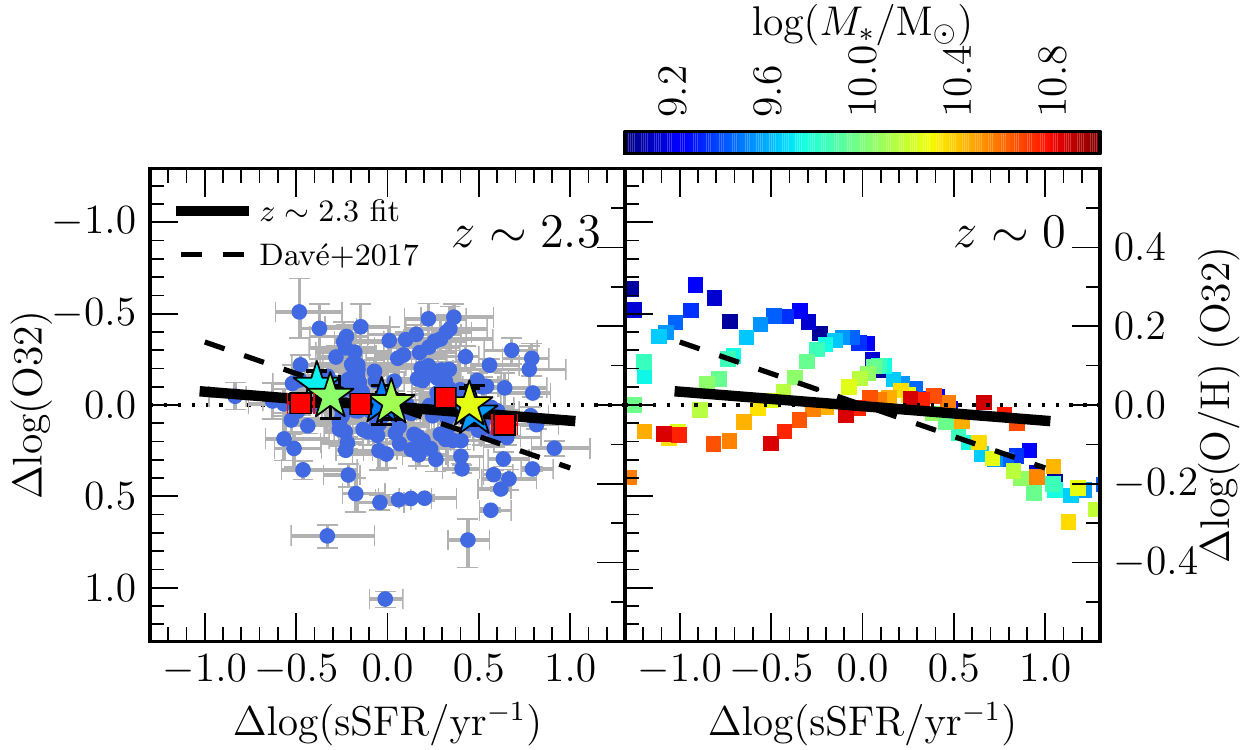}
 \centering
 \caption{Deviation plots for O32 and metallicity estimated using O32, with points and lines the same as
 in Figure~\ref{fig3}.
  In each panel, the left vertical axis shows the scale of the residuals in each line ratio,
 while the right vertical axis displays the scale of the corresponding residuals in metallicity.
  The best-fit slope for the $z\sim2.3$ $\Delta$log(O/H) (O32) vs.~$\Delta$log(sSFR/yr$^{-1}$)
 relation is presented in Table~\ref{tab:slopes}.
}\label{fig4}
\end{figure}

We show the line-ratio and MZR residual plots for O32 in Figure~\ref{fig4}.
  There is no clear trend toward lower metallicity with increasing $\Delta$log(sSFR)
 among the $z\sim2.3$ galaxies when utilizing the O32 indicator.
  There is no statistically significant correlation present (Table~\ref{tab:slopes}).
  In contrast, a clear \mstar-SFR-Z relation is present at $z\sim0$ for O32.
  We discuss the lack of SFR dependence at $z\sim2.3$ for the O32 MZR in Section~\ref{sec:o32}.

\begin{figure*}
 \includegraphics[width=0.67\textwidth]{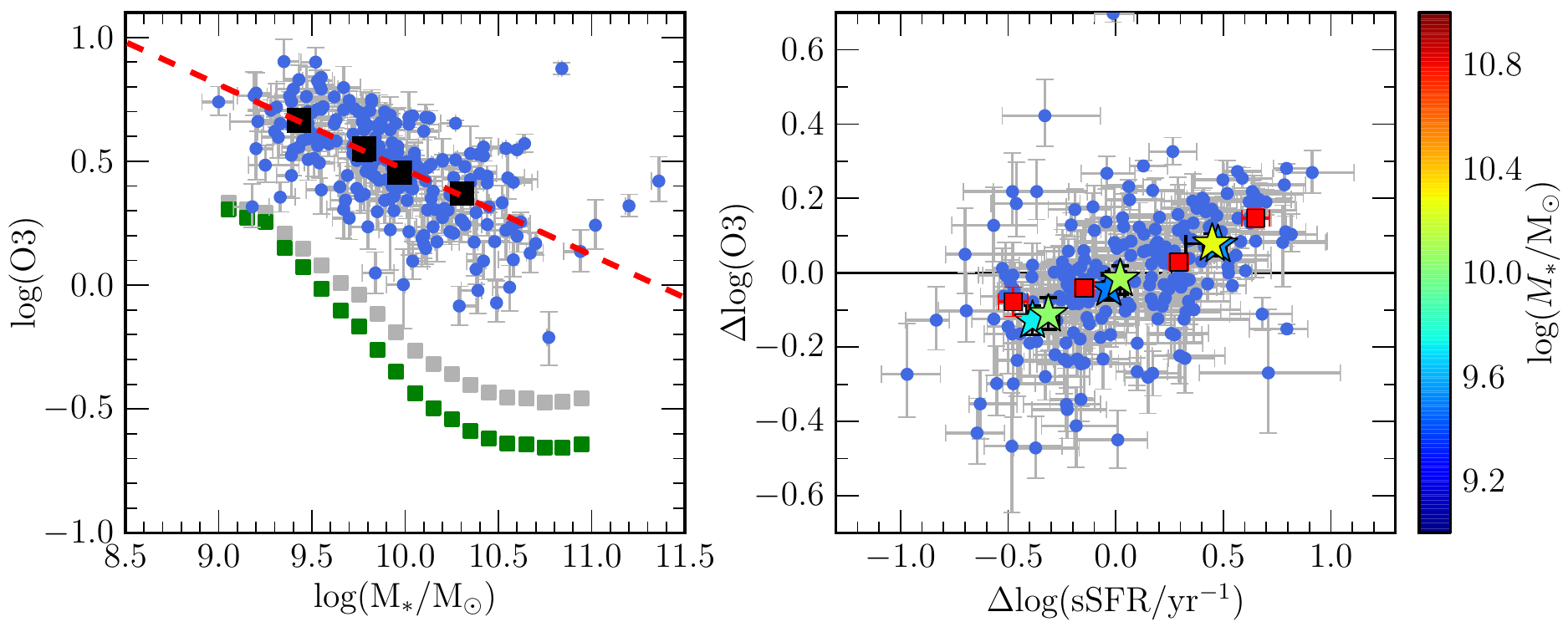}
 \centering
 \caption{\textsc{Left:} The line ratio O3 as a function of \mstar\ for the $z\sim2.3$ and $z\sim0$ samples,
 with points and lines the same as in Figure~\ref{fig2}.
  \textsc{Right:} The deviation plot of $\Delta$log(O3) vs.~$\Delta$log(sSFR/yr$^{-1}$) for $z\sim2.3$
 galaxies, with points the same as in the middle column of Figure~\ref{fig3}.
}\label{fig5}
\end{figure*}

The strong-line ratio O3 is sensitive to both ionization parameter and metallicity.
  As such, trends indicative of
 a \mstar-SFR-Z relation should be present in the O3 residuals as well. 
  We show O3 as a function of \mstar\ in the left panel of Figure~\ref{fig5}.  The best-fit coefficients
 are given in Table~\ref{tab:bestfits}.  The right panel of Figure~\ref{fig5} displays the O3 residuals
 as a function of sSFR residuals for individual $z\sim2.3$ galaxies and the \mstar-$\Delta$sSFR stacks.
  A clear correlation is present, with Spearman correlation coefficient $r_{\text{s}}=0.47$ and probability of
 being drawn from an uncorrelated distribution of $1.6\times10^{-13}$.
  The correlation between $\Delta$log(O3) and $\Delta$log(sSFR) is consistent with a decrease in metallicity
 (increase in O3) as sSFR increases at fixed \mstar.  This result agrees with what is found
 using O3N2, N2, and N2O2.

\begin{figure*}
 \includegraphics[width=0.67\textwidth]{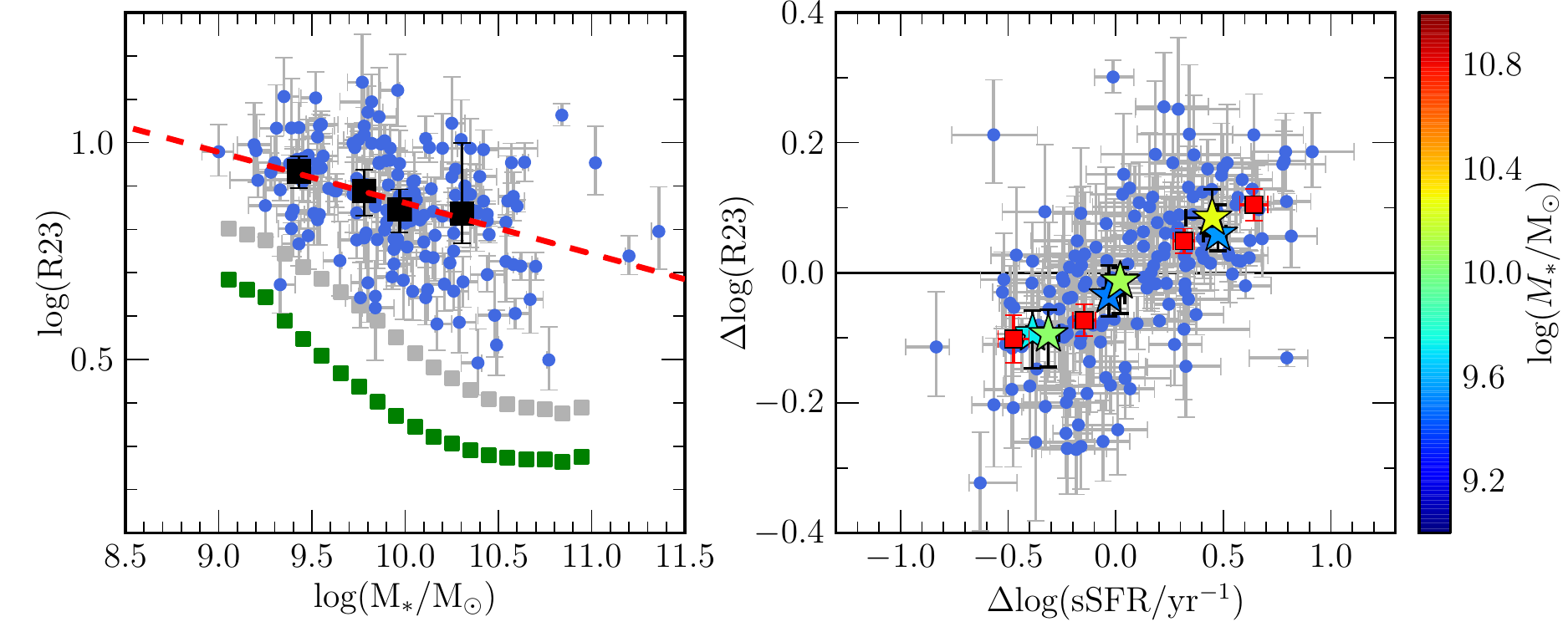}
 \centering
 \caption{\textsc{Left:} The line ratio R23 as a function of \mstar\ for the $z\sim2.3$ and $z\sim0$ samples,
 with points and lines the same as in Figure~\ref{fig2}.
  \textsc{Right:} The deviation plot of $\Delta$log(R23) vs.~$\Delta$log(sSFR/yr$^{-1}$) for $z\sim2.3$
 galaxies, with points the same as in the middle column of Figure~\ref{fig3}.
}\label{fig6}
\end{figure*}

The strong-line ratio R23 as a function of \mstar\ and the $\Delta$log(R23) vs.~$\Delta$log(sSFR)
 deviation plot are presented in Figure~\ref{fig6}.
  R23 is primarily sensitive to O/H because it is a ratio of both low and high ionization state oxygen
 lines to a hydrogen Balmer line, but has significant ionization parameter dependence as well \citep{kew02}.
  R23 is not a strong function of \mstar\ for the $z\sim2.3$ sample, which is anticipated since the R23 values lie close
 to the turnover regime in the R23-O/H relation, as described in Section~\ref{sec:metallicities}.
  However, R23 does increase slightly with decreasing \mstar, suggesting that our
 sample mostly lies on the upper metal-rich R23 branch where R23 increases with decreasing O/H.
  In this case, a positive correlation between $\Delta$log(R23) vs.~$\Delta$log(sSFR) is expected if a
 \mstar-SFR-Z relation is present.  This positive correlation is observed in the right panel of Fig.~\ref{fig6}
 with high statistical significance ($r_{\text{s}}=0.60$, $p\mbox{-value}=7.3\times10^{-18}$),
 in agreement with the trends observed using O3N2, N2, N2O2, and O3.

One important question is whether the SFR dependence we find in the $z\sim2.3$ MZR
 is a result of selection effects.
  We investigated the effect of requiring H$\beta$ S/N$\geq$3 on these results
 by repeating the analysis to search for a \mstar-SFR-Z relation at $z\sim2.3$ without this requirement. 
  Because we cannot reliably correct for reddening without detections of both H$\beta$ and H$\alpha$,
 we utilized SFRs estimated from FAST SED fitting instead of dust-corrected H$\alpha$ luminosities.
  Furthermore, this test could only be performed for the portions of the analysis that
 involve line ratios that do not require reddening correction (i.e., O3, N2, and O3N2).  
  We found similar results based on these line ratios, although the trends were somewhat weaker
 for a few reasons.  First, there is significant scatter in the correlation between SFR(SED) and SFR(H$\alpha$)
 \citep{red15}.  Second, SFR(H$\alpha$) traces star formation on $<$10~Myr timescales while SFR(SED) probes
 longer timescales of $\sim$100~Myr.  It is thus expected that signatures of a \mstar-SFR-Z relation are
 weaker when relying on SFR(SED) since the \mstar-SFR-Z relation is thought to be driven by a reaction of SFR and
 metallicity on short timescales to recent accretion of metal-poor gas \citep{man10,dav12}.
  Finally, a smaller dynamic range of SFR at fixed \mstar\ is probed in this test because \mstar\ and SFR(SED)
 are not independent of one another, both being estimated by FAST SED fitting.  This covariance artificially
 tightens the \mstar-(s)SFR relation, making it more difficult to resolve SFR dependence of metallicity at
 fixed \mstar.  Because our results persist when utilizing SFR(SED) without the H$\beta$ detection criterion,
 we conclude that requiring H$\beta$ S/N$\geq$3 does not strongly bias our results.

\subsection{Do $z\sim2.3$ galaxies lie on the $z\sim0$ FMR?}

Having established the existence of a \mstar-SFR-Z relation for $z\sim2.3$ star-forming
 galaxies, we next address the question of whether or not these galaxies lie on
 the $z\sim0$ \mstar-SFR-Z relation.  In other words, we test whether or a FMR exists
 that extends out to $z\sim2.3$.
  Many studies have attempted to address
 this question by extrapolating parameterizations of the $z\sim0$ FMR \citep[e.g.,][]{man10,lar10}
 to the high-SFR and high-sSFR regime occupied by high-redshift galaxies
 \citep[e.g.,][]{wuy12,wuy14,chr12,bel13,hen13b,cul14,yab14,zah14b,kas17}.
  Such extrapolations may not be representative of the true $z\sim0$ \mstar-SFR-Z relation
 because a parametric form must be assumed that may not represent the underlying physical
 relation, and
 the small fraction of $z\sim0$ objects with SFRs similar to those of high-redshift galaxies
 will not carry much weight toward the fit,
 allowing the possibility of a poor fit in the high-SFR regime.
  Recent studies have shown the benefit of performing
 non-parametric comparisons instead, relying on the small number of $z\sim0$ objects
 with extreme SFRs to directly compare at fixed \mstar\ and SFR \citep{san15,sal15,ly16}.

\begin{figure*}
 \includegraphics[width=1.\textwidth]{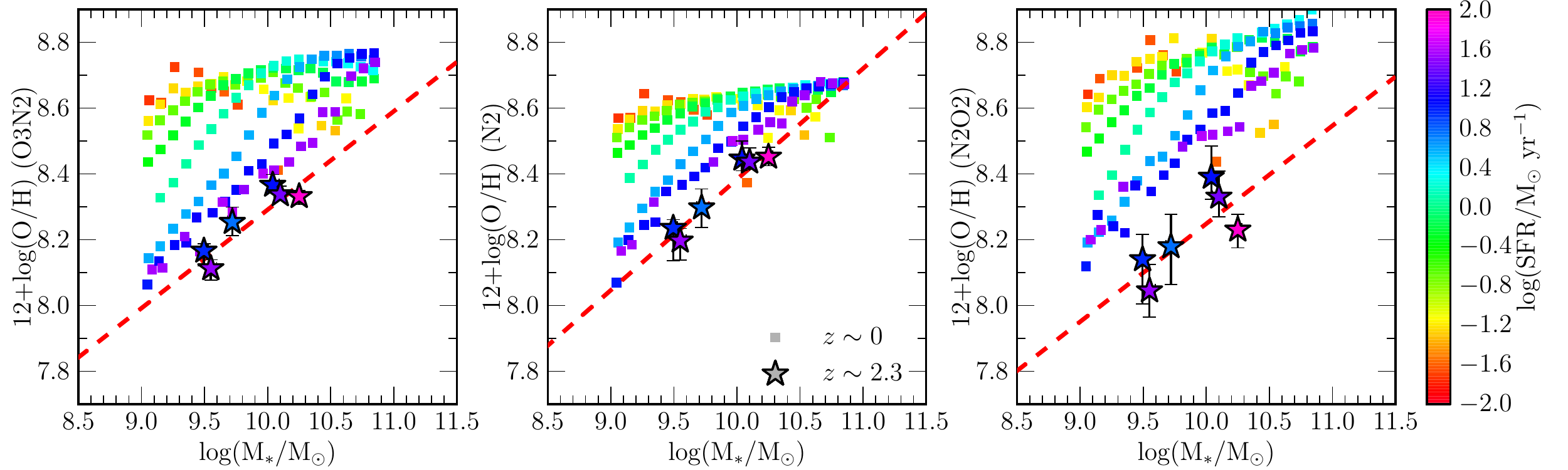}
 \centering
 \caption{The mass-metallicity relation based on O3N2 (left), N2 (middle), and N2O2 (right)
 for $z\sim0$ and $z\sim2.3$ stacks, color-coded by SFR.  Colored squares indicate the $z\sim0$
 \mstar-SFR stacks of \citet{and13}.  Colored stars with error bars show the $z\sim2.3$ \mstar-$\Delta$sSFR stacks.
  Both samples are color-coded by SFR on the same scale.  The red dashed line denotes the
 best-fit $z\sim2.3$ MZR for each line ratio, given in Table~\ref{tab:bestfits}.
}\label{fig7}
\end{figure*}

We directly compare the position of the $z\sim2.3$ \mstar-$\Delta$sSFR stacks and the $z\sim0$
 \mstar-SFR stacks in the mass-metallicity plane in Figure~\ref{fig7}, with
 metallicities based on the O3N2, N2, and N2O2 indicators.
  Both high- and low-redshift stacks are color-coded by SFR on the same scale
 so that a direct comparison of metallicity at fixed \mstar\ and SFR is possible.
  The highest-SFR $z\sim2.3$ stack, with log(M$_*$/M$_{\odot})=10.2$ and
 log(SFR/M$_{\odot}$~yr$^{-1})\approx2.0$, does not have any local analogue in the AM13 \mstar-SFR
 sample.  However, all other $z\sim2.3$ stacks have analogous $z\sim0$ counterparts
 matched in \mstar\ and SFR.  In the O3N2, N2, and N2O2 \mstar-Z planes,
 $z\sim2.3$ galaxies display metallicities that are systematically lower than their $z\sim0$
 counterparts by $\sim0.1$~dex at fixed \mstar\ and SFR.  This offset is consistent across
 a range of \mstar\ and $\Delta$log(sSFR) using three different metallicity indicators.
  We found a similar offset using only O3N2 and N2 with a smaller $z\sim2.3$ MOSDEF sample in
 \citet{san15}.
  We conclude that there is not a FMR
 that can simultaneously match the properties of star-forming galaxies from $z\sim0$ out to $z\sim2.3$.

We note that the presence of a $z\sim2.3$ \mstar-SFR-Z relation can be seen in Figure~\ref{fig7} in the
 \mstar-$\Delta$sSFR stacks, with the lowest-SFR bins at a given \mstar\ falling above the mean $z\sim2.3$
 relation and the highest-SFR bins falling below.  This trend is most evident in the
 O3N2 and N2O2 mass-metallicity planes (Fig.~\ref{fig7}) for the $z\sim2.3$ stacks with
 log(M$_*$/M$_{\odot})=10.0-10.3$, as shown by a progression from higher to lower O/H as SFR increases,
 and the symbol color changes from blue to purple to pink.
  The presence of such SFR gradients perpendicular to
 the mean $z\sim2.3$ MZR is a manifestation of the sSFR dependence shown in Figure~\ref{fig3}.

\begin{figure*}
 \includegraphics[width=1.\textwidth]{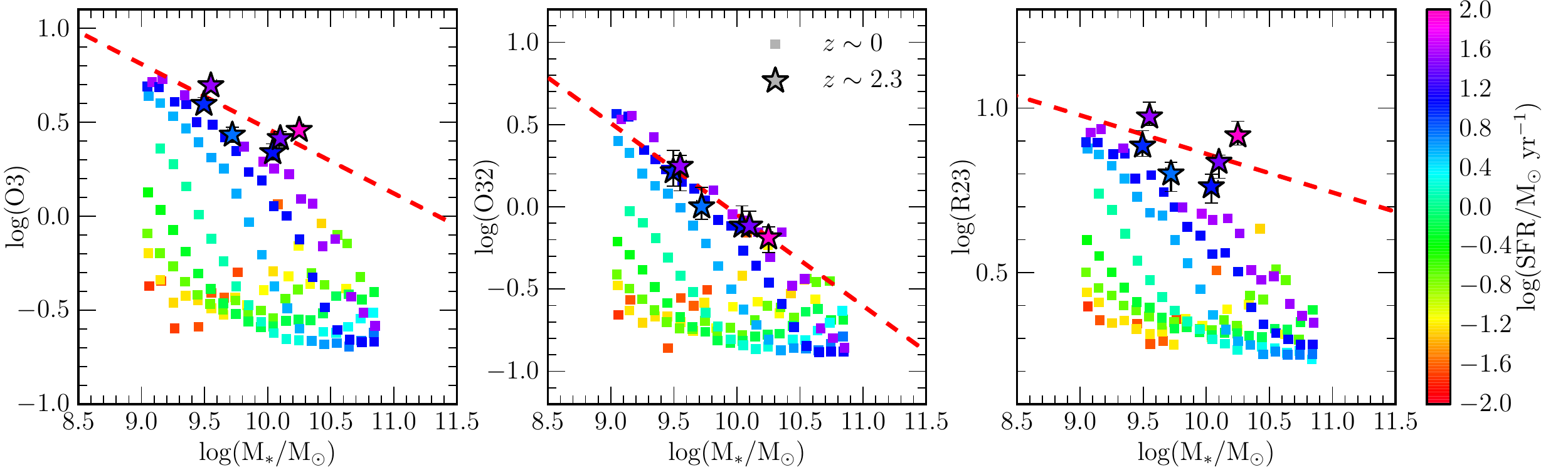}
 \centering
 \caption{The line ratios O3 (left), O32 (middle), and R23 (right) as a function of \mstar.  Points and lines
 are the same as in Figure~\ref{fig7}.
}\label{fig8}
\end{figure*}

We directly compare the strong-line ratios O3, O32, and R23 at fixed \mstar and SFR
 for $z\sim0$ and $z\sim2.3$ stacks in Figure~\ref{fig8}.
  The $z\sim2.3$ stacks display higher O3 and R23 than $z\sim0$ stacks at fixed \mstar\ and SFR,
 suggesting higher excitation and lower metallicity at fixed \mstar\ and SFR at $z\sim2.3$.
  This result is consistent with what was found using O3N2, N2, and N2O2.
  The $z\sim2.3$ \mstar-$\Delta$sSFR stacks also display separation perpendicular to the mean O3-\mstar\
 and R23-\mstar\ relations
 as a function of SFR, consistent with the existence of a \mstar-SFR-Z relation at $z\sim2.3$.
  The O32 values of the $z\sim2.3$ stacks are not obviously offset from the $z\sim0$ stacks
 matched in \mstar\ and SFR.  Once again, results based upon O32 are inconsistent with those
 from O3N2, N2, N2O2, O3, and R23.  We discuss the differences in results based on O32 in Section~\ref{sec:o32}.

\section{Discussion}\label{sec4}

\subsection{Potential evolution in metallicity calibrations and ionized gas physical conditions}\label{sec:evolution}

We have clearly demonstrated the existence of a \mstar-SFR-Z relation at $z\sim2.3$
 based on multiple emission-line ratios (O3N2, N2, N2O2, O3, and R23) and metallicities inferred
 using $z\sim0$ calibrations.  We furthermore showed that $z\sim2.3$ galaxies are offset
 $\sim0.1$~dex lower in metallicity from $z\sim0$ galaxies at fixed \mstar\ and SFR,
 arguing against the existence of a redshift invariant \mstar-SFR-Z relation.
  We now consider whether the appearance of a \mstar-SFR-Z relation at $z\sim2.3$
 and its evolution with respect to $z\sim0$ could be produced by other effects in the
 absence of metallicity variations.
  Recent work has suggested that the physical conditions of ionized gas in star-forming
 regions evolve with redshift \citep[e.g.,][]{ste14,ste16,sha15,san16a,str17,kas17}.
  The physical conditions of interest are the ionization parameter, N/O abundance ratio,
 electron density, and the shape of the ionizing spectrum.  Changes in these properties
 lead to changes in emission-line ratios at fixed nebular oxygen abundance \citep{kew13}.
  The ionization parameter, N/O
 ratio, and shape of the ionizing spectrum must be considered at fixed metallicity because
 of correlations between these properties and the nebular and stellar metallicity.

The only evolving physical condition for which there is a consensus is the electron density.
  The typical electron density in star-forming regions increases with redshift,
 and is an order of magnitude higher at $z\sim2$ compared to $z\sim0$ \citep[e.g.,][]{ste14,san16a,kas17}.
  The typical densities at $z\sim0$ and $z\sim2.3$ (25 and 250~cm$^{-3}$, respectively) are both well
 below the density where strong-line ratios become strongly affected \citep[$\sim$1000~cm$^{-3}$;][]{ste14}.
  Thus, evolution in electron density has a negligible impact on our results.

There is not a consensus about the evolution (or lack thereof) of N/O, ionization parameter, and
 hardness of the ionizing spectrum at fixed metallicity.  While this ambiguity makes it difficult to assess
 the impact of such evolution on metallicity studies, the MOSDEF sample crucially provides access to
 multiple strong-line ratios, enabling us to determine if evolution in a particular property leads
 to false conclusions.  It has been suggested that high-redshift galaxies have elevated N/O at fixed
 O/H compared to $z\sim0$ galaxies, explaining observed evolution in line-ratio diagrams involving nitrogen
 \citep{mas14,ste14,sha15,san16a}.  The elevated N/O could be caused by an unusually high occurrence rate
 of Wolf-Rayet stars \citep{mas14} or a dilution of O/H by pristine gas inflows while N/O remains roughly
 constant \citep{kop05,amo10,san16a}.
  Other studies have argued for a harder ionizing spectrum at fixed nebular abundance in high-redshift
 star-forming regions \citep{ste16,str17}.
  Some have also suggested a higher ionization parameter at fixed metallicity as sSFR increases,
 leading to an elevated ionization parameter in high-redshift galaxies due to their larger sSFR
 than $z\sim0$ galaxies on average \citep{kew13,kew13t,kew15,kew16,kas17}.
  In what follows, we consider the possible effects of these proposed scenarios on our results.

\subsubsection{Nitrogen-to-oxygen ratio}

We observe the $z\sim2.3$ \mstar-SFR-Z relation using the O3N2, N2, and N2O2 metallicity indicators
 (Fig.~\ref{fig3}), all of which involve nitrogen.
  Even if high-redshift galaxies have elevated N/O at fixed O/H, such an effect could only lead to a
 false inference of a \mstar-SFR-Z relation at $z\sim2.3$ if N/O depends on SFR and sSFR at fixed \mstar.
  We observe higher O3N2, lower N2, and lower N2O2 with increasing sSFR at fixed \mstar.
  For these trends to be introduced by variations in nitrogen abundance in the absence of
 metallicity variation, N/O would have to decrease with increasing sSFR at fixed O/H.
  Such a relation between N/O and sSFR at fixed O/H is inconsistent with the Wolf-Rayet
 scenario because it requires that high-sSFR environments underproduce Wolf-Rayet stars
 and thus have lower N/O values at fixed O/H.  In contrast, evolutionary scenario proposed by \citet{mas14}
 predicts that high-sSFR environments \textit{overproduce} Wolf-Rayet stars, leading to elevated
 N/O at fixed O/H in high-redshift galaxies.
  Accordingly, this scenario cannot reproduce our results on the \mstar-SFR-Z relation involving N-based indicators.

In addition to a $z\sim2.3$ \mstar-SFR-Z relation, we found that $z\sim2.3$ galaxies are offset $\sim0.1$~dex
 lower in metallicity at fixed \mstar\ and SFR compared to the $z\sim0$ sample (Fig.~\ref{fig7}).
  If N/O is higher at fixed O/H at high redshifts, then high-redshift metallicities based on O3N2, N2, and N2O2
 would be overestimated.  Consequently, the $\sim0.1$~dex metallicity offset would be underestimated and
 our conclusion that the \mstar-SFR-Z relation evolves with redshift from $z\sim0$ to $z\sim2.3$
 still holds, as argued in \citet{san15}.
  At fixed \mstar, we observe an offset toward lower N2 and N2O2, and higher O3N2,
 for the $z\sim2.3$ sample.
  For this offset to be the product of N/O evolution at fixed O/H,
 high-redshift galaxies would need to have lower N/O at fixed O/H compared to $z\sim0$.
  Such a scenario would shift high-redshift galaxies toward lower N2 at fixed O/H.
  In fact, high-redshift galaxies are observed to be offset significantly \textit{higher} N2
 at fixed O3 compared to the $z\sim0$ star-forming population in the
 N2 vs. O3 diagram \citep[e.g.,][]{sha05,sha15,ste14,san16a,str17}.
  It is thus implausible that evolution toward lower N/O at fixed O/H drives the redshift evolution of O3N2, N2,
 and N2O2 at fixed \mstar\ and SFR.

It is important to note that both the $z\sim2.3$ \mstar-SFR-Z relation and the offset toward lower O/H
 at fixed \mstar\ and SFR compared to $z\sim0$ are also observed using O3 and R23, which have no dependence on variations
 in N/O at fixed O/H (Figs.~\ref{fig5},~\ref{fig7} and~\ref{fig8}, left and right panels).
  Simultaneously obtaining consistent results
 in N-based and O-based metallicity-sensitive line ratios strongly suggests that possible evolution in N/O
 at fixed O/H does not lead us to false conclusions regarding the $z\sim2.3$ \mstar-SFR-Z relation,
 demonstrating the power of using multiple line ratios to study the evolution of metallicity scaling relations.

\subsubsection{Ionization parameter and hardness of the ionizing spectrum}

\citet{ste16} suggested that the ionizing spectrum is harder at fixed nebular metallicity at
 high redshifts because of low Fe/O (and therefore low Fe/H at fixed O/H)
 driven by the young ages of high-redshift stellar populations
 \citep[see also][]{str17}.  This deficiency of Fe relative to O is thought to be a result of the time-delayed nature
 of Fe enrichment from Type Ia supernovae compared to the prompt enrichment of O from
 Type II supernovae.
  In a harder ionizing spectrum, the increased abundance of high-energy ultraviolet photons
 relative to those at lower energy leads
 to larger fractions of ions in high ionization states (i.e., [O\iii]) relative to low-ionization
 states (i.e., [O\ii]).  Thus a harder ionizing spectrum at fixed metallicity would result in higher
 excitation-sensitive line ratios at fixed O/H, including O3 and O3N2.  The scenario in which the ionization
 parameter is higher at fixed O/H due to higher sSFR or more concentrated star-formation at high-redshift
 similarly predicts higher O3 and O3N2 at fixed O/H \citep{kew13,kew16,kas17}.
  However, the two scenarios predict different changes in N2, with a harder ionizing spectrum
 increasing N2 at fixed O/H, while a higher ionization parameter decreases N2 at fixed O/H.

In the scenario proposed by \citet{ste16}, galaxies with the highest sSFR should also have the youngest
 stellar populations in which Fe/H is lowest.  This low Fe abundance in turn
 produces a harder ionizing spectrum at fixed O/H, such that the highest-sSFR galaxies should have
 the hardest ionizing spectra at fixed \mstar.
  If this scenario holds, then it could introduce the observed trends in
$\Delta$log(O3) and $\Delta$log(O3N2) as a function of $\Delta$log(sSFR) in the absence
 of a \mstar-SFR-Z relation.
  However, if a harder ionizing spectrum at fixed oxygen abundance is present, then N2 should
 increase with increasing sSFR at fixed \mstar.  We observe the opposite trend.
  Additionally, this scenario predicts that $z\sim2.3$ galaxies should have higher O3N2, O3,
 and N2 than $z\sim0$ galaxies at fixed O/H.  We observe higher O3N2 and O3, but lower N2
 at $z\sim2.3$ compared to $z\sim0$ galaxies at fixed \mstar\ and SFR.
  Therefore, a harder ionizing spectrum at fixed metallicity driven by Fe/O variations
 cannot simultaneously reproduce the observed trends in O3N2, O3, and N2, and if present would not
 lead us to falsely infer the presence of both a $z\sim2.3$ \mstar-SFR-Z relation and
 an evolving \mstar-SFR-Z relation.

Since the evolving ionization parameter scenario suggests that ionization parameter increases at fixed O/H
 with increasing sSFR, it is straighforward to understand how the trends of $\Delta$log(O3),
 $\Delta$log(O3N2), and $\Delta$log(N2) vs. $\Delta$log(sSFR) could appear in the absence of
 metallicity variation.
  A higher ionization parameter at fixed O/H at high redshift could also introduce the offsets
 observed in O3N2, N2, and O3 between $z\sim0$ and $z\sim2.3$ galaxies at fixed \mstar\ and SFR.
  It is thus possible that an increase in ionization parameter at fixed O/H that is dependent on
 sSFR could reproduce the observed trends in O3N2, N2, and O3 at $z\sim2.3$.

However, N2O2 would not be significantly affected by changes in the ionization parameter,
 since it is a ratio of collisionally excited lines of two low-ionization species with similar ionization
 energies \citep{kew02}.  The observation of the $z\sim2.3$ \mstar-SFR-Z relation and the metallicity offset
 between $z\sim0$ and $z\sim2.3$ galaxies at fixed \mstar\ and SFR based on N2O2 suggests that the
 trends in O3, O3N2, and N2 are not purely driven by changes in ionization parameter at fixed metallicity.
  We conclude that our evidence for the existence of a $z\sim2.3$
 \mstar-SFR-Z relation is not introduced by changes in ionization parameter at fixed O/H as a function of sSFR,
 and that the inferred evolution of the \mstar-SFR-Z relation from $z\sim0$ to $z\sim2.3$ is also not a consequence
 of such evolution.

Collectively, our analyses of potential biases introduced by evolving N/O, ionization parameter, and ionizing spectrum
 at fixed metallicity suggest that the observed trends in O3N2, N2, N2O2, and O3 are primarily driven
 by metallicity variations, and that the $z\sim2.3$ \mstar-SFR-Z relation and evolution in O/H at fixed
 \mstar\ and SFR are real.
  We note that the above discussion does not preclude evolution in gas physical conditions with redshift,
 but instead shows that the examined evolutionary scenarios cannot simultaneously produce our results
 over the range of emission-line ratios used here in the absence of metallicity variation.
  We have considered evolution in each property separately, but it may be the case that a combination of the proposed
 evolutionary scenarios is required to explain high-redshift observations (e.g., both a harder ionizing spectrum
 and higher N/O at fixed O/H).  In that case, a more careful analysis accounting for the magnitude of
 shifts in line ratios from changes in each property would be needed.  However, our results are consistent with
 a change in metallicity with SFR at fixed \mstar\ being the primary driver of the observed trends in
 O3, O3N2, N2, and N2O2.  Evolution in other gas properties likely has a secondary effect on the strength
 of these trends.

\subsection{Implications of the evolution of the \mstar-SFR-Z relation}

The existence of a \mstar-SFR-Z relation at $z\sim2.3$ confirms that the current theoretical
 framework for galaxy growth through the interplay of inflows, outflows, and star formation
 is applicable at high redshifts.  A \mstar-SFR-Z relation is predicted to exist
 at high redshifts in analytical chemical evolution models \citep{fin08,dav12,lil13} as well
 as cosmological hydrodynamical simulations including chemical evolution \citep{ma16,dav17,der17}.
  There are both observational evidence and theoretical predictions that the \mstar-SFR-Z relation
 is a manifestation of a more fundamental relation with gas mass, where the SFR is modulated by
 the amount of gas in a galaxy \citep{hug13,bot13,bot16b,bot16a,zah14a,ma16,dav17}.
  Confirmation of a \mstar-SFR-Z
 relation at $z\sim2.3$ suggests that galaxy chemical evolution is linked to gas content
 at high redshifts as well.  The increasing number of observations of the cold gas content of
 high-redshift galaxies \citep[e.g.,][]{tac13} will allow for investigations of the relation between
 \mstar, SFR, Z, and gas fraction at high redshifts.

By comparing residuals in metallicity and sSFR around the MZR and mean \mstar-sSFR relation,
 we have quantified the strength of the SFR dependence of the \mstar-SFR-Z relation
 at both $z\sim0$ and $z\sim2.3$.
  Current cosmological hydrodynamical simulations predict that the strength of
 the SFR dependence does not change with redshift \citep{dav17,der17}.
  Using the \textsc{mufasa} simulations \citep{dav16}, \citet{dav17} predicted the
 slope of the $\Delta$log(O/H) vs. $\Delta$log(sSFR) relation, finding the preferred slope to
 be $-0.16$ and independent of redshift.  We have quantified this measure of the \mstar-SFR-Z
 relation for our $z\sim2.3$ sample, providing best-fit slopes based on the O3N2, N2, and N2O2
 metallicity indicators in Table~\ref{tab:slopes}.

We compare our observed $z\sim2.3$
 $\Delta$log(O/H) vs. $\Delta$log(sSFR) relations to the prediction of \citet{dav17} in the
 middle column of Figure~\ref{fig3}, and perform the same comparison for the $z\sim0$
 sample in the right column.  In each panel, the solid black line is the best-fit $z\sim2.3$ relation
 and the dashed black line shows the prediction of \citet{dav17}.
  We find weaker SFR dependence at $z\sim2.3$ than predicted by \citet{dav17} based on O3N2 and N2,
 and stronger SFR dependence when using N2O2 to estimate metallicites.  The O3N2, N2, and N2O2
 best-fit slopes all agree with the \citet{dav17} prediction within 2$\sigma$.
  Taking the uncertainty-weighted average of the three $z\sim2.3$ slopes yields a slope of $-0.14\pm0.022$,
 which is within 1$\sigma$ of a slope of $-0.16$.  Our results are thus consistent with
 the predictions of the \textsc{mufasa} simulations, although both the measurement and
 systematic uncertainties remain large.

The SFR dependence of the $z\sim2.3$ sample displays a similar strength to that of $z\sim0$
 galaxies with log($M_*$/M$_{\odot})<10.0$ and $\Delta$log(sSFR/yr$^{-1})>0$ when using
 the N2 and N2O2 indicators (Fig.~\ref{fig3}).  When using the O3N2 indicator, the $z\sim0$ relation appears to be
 somewhat stronger than at $z\sim2.3$.  Overall, the observations do not suggest large changes in the
 strength of the SFR dependence of metallicity at fixed \mstar\ and SFR over the past 10.5 Gyr.
  We caution that a detailed comparison of the strength of the $z\sim0$ and $z\sim2.3$
 \mstar-SFR-Z relation requires a more complete understanding of the redshift evolution of
 gas physical conditions and the corresponding effects on metallicity indicators.
  Recalibration of strong-line ratio metallicity relations at high redshift using
 electron-temperature-based direct metallicities provides a promising avenue to
 eliminate these systematics \citep[e.g.,][]{jon15,san16b}.

The $\sim0.1$~dex offset toward lower metallicity at fixed \mstar\ and SFR observed
 from $z\sim0$ to $z\sim2.3$ demonstrates that the \mstar-SFR-Z relation is not redshift invariant.
  While this shift in metallicity is fairly small, a systematic offset is observed across more than an
 order of magnitude in \mstar\ and SFR using four different metallicity-sensitive line ratios
 (O3N2, N2, N2O2, and O3).  Observing the evolution of the \mstar-SFR-Z relation in multiple
 line ratios simultaneously allowed us to show that this metallicity offset cannot be a false inference
 due to potential evolution of metallicity calibrations with redshift.

In \citet{san15}, we found a similar evolution in O/H at fixed \mstar\ and SFR, and speculated
 that such an offset could be introduced if the inflow rate exceeded the sum of the SFR
 and outflow rate such that the gas reservoir grows faster than it can be processed via
 star formation.  While such a gas-accumulation phase is predicted to exist, both models
 and observations suggest that it should only occur at $z>4$ \citep{dav12,pap11}.
  This phase can be made to reach $z\sim2$ in models with no outflows \citep{dav12},
 which are clearly unphysical given the observational constraints on the occurrence of
 galactic winds at high redshifts \citep[e.g.,][]{ste10}.  It is thus unlikely that
 we are observing the buildup of galaxy gas reservoirs.

If the most fundamental relation is between \mstar, metallicity, and gas content and that relation is redshift invariant,
 then an evolving \mstar-SFR-Z relation would be indicative of an evolving relation between SFR and gas mass.
  If a non-evolving relation between \mstar, gas fraction, and metallicity exists, then
 galaxies at $z\sim0$ and $z\sim2.3$ have the same gas mass at fixed \mstar\ and metallicity.
  The observed trend of decreasing metallicity at fixed \mstar\ and SFR with increasing redshift
 could then be understood as decreasing SFR at fixed \mstar\ and gas fraction (i.e., lower star-formation efficiency).
  However, an inferred decrease in $z\sim2$ star-formation efficiency at fixed \mstar\ is in conflict
 with the interpretation of recent observations of cold gas in high-redshift galaxies.
  \citet{sco16,sco17} found that, at fixed \mstar\, star-formation efficiency per unit ISM gas mass \textit{increases}
 with increasing redshift using ALMA observations of luminous star-forming galaxies at $z\sim1-6$.
  Furthermore, molecular gas depletion timescales of galaxies on the mean \mstar-SFR relation are shorter
 at $z\sim1-3$ than at $z\sim0$ \citep{tac13,gen15}, suggesting an increase in star-formation efficiency
 with redshift.  Reconciling these observations with the observed evolution of the \mstar-SFR-Z relation
 requires an evolving relationship among \mstar, gas fraction, and metallicity
 such that metallicity decreases at fixed \mstar\ and gas fraction with increasing redshift.
  Since a decreasing star-formation efficiency with increasing redshift at fixed \mstar\
 conflicts with observational constraints, we must investigate other drivers of \mstar-SFR-Z evolution.

The cause of an evolving \mstar-SFR-Z relation can also be understood through analytical chemical evolution models.
  The gas-regulator model of \citet{lil13} allows for an evolving \mstar-SFR-Z relation
 if the gas consumption timescale and/or the mass-loading factor ($\eta$=outflow rate/SFR)
 at fixed \mstar\ change with redshift.
  In this context, an evolving relation toward lower metallicity at fixed \mstar\ and SFR
 with increasing redshift suggests that, at fixed \mstar, either the gas consumption timescale is longer
 (i.e., star-formation efficiency is lower) at higher redshifts or $\eta$ is higher
 at earlier times.
  The equilibrium model of \citet{dav12} similarly predicts lower O/H at fixed \mstar\ and SFR
 at high redshifts if $\eta$ increases at fixed \mstar\ with increasing redshift.
  Although a lower star-formation efficiency at higher redshift appears to be in conflict with the
 observations detailed above, a larger typical high-redshift mass-loading factor is at least broadly
 consistent with observational constraints over a wide range of \mstar\ and redshift
 \citep[e.g.,][]{pet02,rup05,wei09,ste10,hec15,chi17}.
  We do note, however, that estimates of $\eta$ carry large systematic uncertainties.

In both the equilibrium model of \citet{dav12} and the gas-regulator model of \citet{lil13},
 the ISM metallicity at fixed \mstar\ and SFR decreases if the metallicity of the inflowing gas decreases.
  Although accreted gas is relatively unenriched compared to the ISM, it is not metal-free.
  Numerical simulations have shown that accreted gas is a mix of relatively pristine gas from the IGM
 and enriched gas that was ejected in past outflows \citep[e.g.,][]{opp10,seg16,ang17}.
  Recycled gas can even become the dominant source of fuel for star formation at $z<1$ \citep{opp10}.
  Thus evolution toward lower metallicity at fixed \mstar\ and SFR at $z\sim2.3$ could also be
 explained by a decrease in the metallicity of infalling gas at high redshift caused by a lower importance
 of recycled gas accretion relative to pristine accretion from the IGM.  Numerical simulations agree with this
 scenario.  Indeed, in the models of \citet{dav11}, an increase in the metallicity of infalling gas relative to that of
 the ISM was the primary driver of the increase in metallicity at fixed \mstar\ with decreasing redshift.
  \citet{ang17} found that the relative importance of recycled gas to total accretion increases with
 decreasing redshift in the FIRE zoom-in simulations
 and additionally suggest that transfer of metals between galaxies via galactic winds occurs at $z=0$,
 further increasing the metallicity of infalling gas at $z\sim0$ compared to that at $z\sim2$.

We conclude that the offset toward lower O/H at fixed \mstar\ and SFR with increasing redshift
 is most likely caused by an increase in the mass-loading factor at fixed \mstar\ from $z\sim0$ to $z\sim2$
 and a decrease in the metallicity of infalling gas due to a higher relative importance of pristine accretion
 from the IGM at $z\sim2$.

\subsection{The inconsistency of results based on O32}\label{sec:o32}

While results based on O3N2, N2, N2O2, O3, and R23 are all consistent with respect to both
 the existence of a $z\sim2.3$ \mstar-SFR-Z relation and the $\sim0.1$~dex offset toward
 lower metallicity at fixed \mstar\ and SFR, performing the same analysis using O32
 yields different results.  The $\Delta$log(O32) vs. $\Delta$log(sSFR) diagram shows
 no evidence for SFR dependence of O/H at fixed \mstar\ (Fig.~\ref{fig4}).  Furthermore,
 there is little if any evolution in O32 at fixed \mstar\ and SFR from $z\sim0$ to $z\sim2.3$
 (Fig.~\ref{fig8}, middle panel).
  The disagreement between results based on O32 compared to those based on the other line ratios
 requires closer examination.

Since the emission lines in O32 are widely separated in wavelength, it is natural to suspect
 that incorrect reddening corrections may give rise to discrepant results based on O32.
  If there are large
 galaxy-to-galaxy deviations from the assumed \citet{car89} attenuation curve, a trend in O32
 with SFR at fixed \mstar\ could be washed out.  
  However, systematic effects associated with galaxy-to-galaxy variations in the attenuation curve
 do not appear to have a significant effect on the O32 results, since we see strong trends in N2O2 and R23
 (which also require a reddening correction) with SFR at fixed \mstar.

  SFR dependence of O32 at fixed \mstar\ could also be hidden if our assumed dust-correction recipe
 overestimates E(B-V)$_{\text{gas}}$ for galaxies lying above the mean $z\sim2.3$ SFR-\mstar\ relation
 and/or underestimates E(B-V)$_{\text{gas}}$ in the opposite regime.
   In this case, O32 values at low $\Delta$log(sSFR)
 would be overestimated while the opposite would occur at high $\Delta$log(sSFR), masking the presence of
 a \mstar-SFR-Z relation. 
  Interestingly, such a scenario would \textit{steepen} $\Delta$log(R23) vs.~$\Delta$log(sSFR) and
 $\Delta$log(N2O2) vs.~$\Delta$log(sSFR) beyond what is expected based on the \mstar-SFR-Z relation because
 R23 increases and N2O2 decreases with increasing SFR at fixed \mstar\ (Figs.~\ref{fig3} and~\ref{fig6}).
  Indeed, we find that the slope of $\Delta$log(O/H) vs.~$\Delta$log(sSFR) is steepest
 when metallicities are based on N2O2, and that $\Delta$log(R23) vs.~$\Delta$log(sSFR) displays
 the strongest correlation of all line ratios considered here.
  It is therefore possible that our dust-correction recipe
 leads to either underestimated E(B-V)$_{\text{gas}}$ for $z\sim2.3$ galaxies below the SFR-\mstar\ relation,
 overestimated E(B-V)$_{\text{gas}}$ for objects above the SFR-\mstar\ relation, or both.
  If true, a bias in E(B-V)$_{\text{gas}}$ dependent on SFR suggests that the nebular attenuation curve
 systematically changes with SFR at fixed \mstar\ at $z\sim2.3$.
  Studies of the stellar continuum attenuation curve and the differential reddening between the continuum
 and nebular emission lines at high redshifts have similarly suggested a systematic dependence of the shape of the attenuation
 curve on SFR and sSFR \citep{kri13,pri14,red15}.
  We note that the resulting bias would affect the measured strength of the SFR dependence of O/H at fixed
 \mstar\ using O3N2, N2, and O3 as well, since biases in E(B-V)$_{\text{gas}}$ will change $\Delta$log(sSFR).

Biases in reddening corrections may also explain why no evolution in O32 at fixed \mstar\ and SFR
 is observed from $z\sim0$ to $z\sim2.3$.  If E(B-V)$_{\text{gas}}$ is systematically overestimated in the
 $z\sim2.3$ sample, then the high-redshift O32 values would be underestimated and fall closer
 to the $z\sim0$ galaxies, reducing any offset present.  If this is the case, then the offset
 in O/H at fixed \mstar\ and SFR based on N2O2 is \textit{overestimated}.
  In Figure~\ref{fig7}, the metallicity offset at fixed \mstar\ and SFR based on N2O2
 is larger than what is found using O3N2 and N2, consistent with an overestimate of the
 reddening correction at $z\sim2.3$.  It is thus plausible that biases in the dust-correction
 can explain the discrepant O32 results.

Unlike O3, O3N2, N2, N2O2, and R23, O32 directly probes the ionization parameter and is only
 sensitive to metallicity because of the anticorrelation between ionization parameter and metallicity.
  This anticorrelation arises because gas-phase
 metallicity and stellar metallicity are correlated, giving rise to harder ionizing spectra
 and, consequently, higher ionization parameters in lower metallicity star-forming regions
 \citep{dop06a,dop06b}.
  The O32 results could be explained if ionization parameter does not depend on metallicity
 at $z\sim2.3$.  However, such a scenario would weaken trends in O3N2 and O3, both of which are
 strongly sensitive to ionization parameter in addition to metallicity.  We observe
 clear trends in O3N2 and O3 as a function of SFR at fixed \mstar.
  Furthermore, O32 is clearly still related to metallicity given that O32
 is anticorrelated with stellar mass, and a similar MZR is obtained when using O32
 compared to those based on O3N2, N2, and N2O2 (Fig.~\ref{fig2}).
  We conclude that ionization parameter and metallicity are still anticorrelated at $z\sim2.3$,
 though not necessarily in the same form as at $z\sim0$.

Another possibility is that the scatter in the ionization parameter vs. metallicity relation
 is significantly larger at high redshifts than at $z\sim0$.  If this is the case, additional
 scatter would be introduced in the relationship between O32 and metallicity, potentially
 washing out trends in O32 with SFR at fixed \mstar.  This scenario can be tested by
 measuring the scatter in O32 at fixed metallicity at $z\sim2.3$, where metallicity is based on
 temperature-sensitive auroral-line measurements, but a data set on which to perform such
 an analysis does not yet exist.

Based on the information currently available, biases arising from the reddening correction
 are the most likely explanation of the discrepant O32 results.  These biases are such that
 E(B-V)$_{\text{gas}}$ is systematically overestimated at $z\sim2.3$, and the magnitude of this overestimate
 increases with increasing SFR at fixed \mstar.

\section{Summary and conclusions}\label{sec5}

We have investigated the relationship between \mstar, SFR, and gas-phase oxygen abundance
 at high redshifts using a large, representative sample of $z\sim2.3$ star-forming galaxies from the
 MOSDEF survey.  The MOSDEF data set allows us to study this relation using multiple
 metallicity- and excitation-sensitive line ratios (O3N2, N2, N2O2, O32, O3, R23), and
 derive robust dust-corrected SFRs based on H$\alpha$ luminosity.
  We searched for the presence of a \mstar-SFR-Z relation at $z\sim2.3$ by looking for
 correlations between the scatter around the MZR and the mean $z\sim2.3$ \mstar-sSFR
 relation.  We additionally performed a direct comparison of stacks of $z\sim2.3$ and $z\sim0$
 galaxies matched in \mstar\ and SFR to determine whether or not the \mstar-SFR-Z relation
 evolves with redshift.  Our main results are summarized below.

We presented the detection of a \mstar-SFR-Z relation at $z\sim2.3$ based on
 metallicities estimated using three different emission-line ratios (O3N2, N2, and N2O2).
  While it is non-trivial to estimate metallicity using O3 and R23 due to a double-valued nature,
 results based on observed O3 and R23 are consistent.
  This study uniquely performed an analysis of the \mstar-SFR-Z relation using multiple
 emission lines simultaneously, which allowed us to rule out the possibility of evolving
 physical conditions in high-redshift star-forming regions introducing the observed trends
 in the absence of a \mstar-SFR-Z relation.  Our results cannot be reproduced by the
 proposed redshift evolution of electron density, N/O, ionization parameter, or shape of the
 ionizing spectrum at fixed metallicity in the absence of metallicity variations.
  However, our analysis does not confirm or rule out the presence of any of the proposed
 evolutionary scenarios in our high-redshift sample.
  We find that the SFR dependence of the observed $z\sim2.3$ \mstar-SFR-Z relation
 is similar in strength to the predictions from the recent \textsc{mufasa} cosmological
 hydrodynamical simulations of \citet{dav17}.

We show strong evidence that the \mstar-SFR-Z relation is not redshift invariant, but evolves
 with redshift such that $z\sim2.3$ galaxies have $\sim0.1$~dex lower metallicity than
 $z\sim0$ galaxies at fixed \mstar\ and SFR.  Our results are consistent between five different
 line ratios (O3N2, N2, N2O2, O3, and R23).  These results argue against the existence of a
 non-evolving \mstar-SFR-Z relation \citep{man10}.
  The evolving \mstar-SFR-Z relation suggests an evolving relation between \mstar, gas fraction, and metallicity
 based on observations implying higher star-formation efficiency and lower molecular gas depletion times
 at high redshifts \citep{tac13,gen15,sco16,sco17}.
  Furthermore, we found that the trend toward lower O/H at fixed \mstar\ and SFR is likely driven by
 an increase in the mass-loading factor and a decrease in the metallicity of infalling gas with redshift
 at fixed \mstar.  The latter suggests that accretion of recycled gas is less important relative to
 accretion of pristine gas from the IGM at $z\sim2$ than at $z\sim0$.

Despite observing consistent trends using O3N2, N2, N2O2, O3, and R23, we found that results
 based upon O32 do not display the presence of a \mstar-SFR-Z relation or significant
 redshift evolution in metallicity at fixed \mstar\ and SFR.  Given the agreement of results
 based on all other line ratios, we conclude that systematic effects must bias the observed trends in O32.
  A comparison of results based on O32, N2O2, and R23 (all of which require reddening corrections)
 suggests that E(B-V)$_{\text{gas}}$ is overestimated at $z\sim2.3$ when E(B-V)$_{\text{gas}}$
 is determined using H$\alpha$/H$\beta$ and the \citet{car89} Milky Way extinction curve,
 and that the magnitude of this overestimate increases with increasing SFR at fixed \mstar.
  Reevaluation of the nebular attenuation curve may be required at high redshifts.

The existence of a \mstar-SFR-Z relation at $z\sim2.3$ confirms that the theoretical
 framework through which we view galaxy evolution is applicable at high redshifts.
  Interplay among gas inflows, the gas reservoir, star formation, feedback, and
 gas outflows have shaped the growth history of the baryonic content of galaxies since
 at least $z\sim2.3$.
  This analysis shows the promise of statistical high-redshift samples to make higher-order
 tests of cosmological hydrodynamical simulations including chemical evolution.
  Our ability to utilize rich high-redshift spectroscopic data sets such as the MOSDEF
 data set to the fullest requires a complete understanding of the evolution of physical
 conditions in star-forming regions and metallicity calibrations with redshift.
  Until such understanding is attained, detailed quantitative comparisons of high-redshift
 data and chemical evolution models will necessarily need caveats.
  New observational facilities on the horizon such as the James Webb Space Telescope
 (JWST) and Thirty Meter Telescope (TMT) will obtain large samples of temperature-sensitive
 auroral-line measurements for high-redshift galaxies that will revolutionize the determination
 of metallicities at high redshifts.

\acknowledgements We acknowledge support from NSF AAG grants AST-1312780, 1312547, 1312764, and 1313171,
 archival grant AR-13907 provided by NASA through the Space Telescope Science Institute,
 and grant NNX16AF54G from the NASA ADAP program.
  R.L.S. is supported by a UCLA Graduate Division Dissertation Year Fellowship.
  We additionally acknowledge the 3D-HST collaboration
 for providing spectroscopic and photometric catalogs used in the MOSDEF survey.
  We wish to extend special thanks to those of Hawaiian ancestry on
 whose sacred mountain we are privileged to be guests. Without their generous hospitality,
 the work presented herein would not have been possible.

\bibliography{fmrpaper}

\end{document}